\documentclass[a4paper,11pt]{article}
\usepackage{epsfig,epic}
\usepackage{wrapfig}

\newcommand{\resection}[1]{\setcounter{equation}{0}\section{#1}}

\newcommand{\EQ}{\begin{equation}}
\newcommand{\EN}{\end{equation}}
\newcommand{\bea}{\begin{eqnarray}}
\newcommand{\eea}{\end{eqnarray}}

\newcommand{\st}{\stackrel}
\newcommand{\var}{\varepsilon}

\newcommand{\goto}{\rightarrow}
\newcommand{\lab}{\label}

\begin{document}
\setcounter{page}{0}
\topmargin 0pt
\oddsidemargin 5mm
\renewcommand{\thefootnote}{\arabic{footnote}}
\newpage
\setcounter{page}{0}
\begin{titlepage}
\begin{flushright}
SISSA 105/2003/FM
\end{flushright}
\vspace{0.5cm}
\begin{center}
{\large {\bf Integrable field theory and critical phenomena.}}\\ 
{\large {\bf The Ising model in a magnetic field }}\\
\vspace{1.8cm}
{\large Gesualdo Delfino} \\
\vspace{0.5cm}
{\em International School for Advanced Studies (SISSA)}\\
{\em via Beirut 2-4, 34014 Trieste, Italy}\\
{\em INFN sezione di Trieste}\\
{\em E-mail: delfino@sissa.it}\\
\end{center}
\vspace{1.2cm}

\renewcommand{\thefootnote}{\arabic{footnote}}
\setcounter{footnote}{0}

\begin{abstract}
\noindent
The two-dimensional Ising model is the simplest model of statistical 
mechanics exhibiting a second order phase transition. While in absence of 
magnetic field it is known to be solvable on the lattice since Onsager's
work of the forties, exact results for the magnetic case have been missing
until the late eighties, when A.~Zamolodchikov solved the model in a field
at the critical temperature, directly in the scaling limit, within the 
framework of integrable quantum field theory. In this article we review this 
field theoretical approach to the Ising universality class, with particular
attention to the results obtained starting from Zamolodchikov's scattering
solution and to their comparison with the numerical estimates on the lattice.
The topics discussed include scattering theory, form factors, correlation
functions, universal amplitude ratios and perturbations around 
integrable directions. Although we restrict our discussion to the Ising model, 
the emphasis is on the general methods of integrable quantum field theory 
which can be used in the study of all universality classes of critical
behaviour in two dimensions.
\end{abstract}
\end{titlepage}

\newpage

\resection{Introduction}
A lattice system close to a second order phase transition point exhibits a 
number of features which do not depend on the specific microscopic
realisation and coincide with those of all systems sharing the same essential 
symmetry properties in the given spatial dimension. In principle, the continous
field theoretical description of the scaling region provides the most natural 
theoretical framework for the quantitative study of these universal features; 
in practice, however, the need of non-perturbative methods seriously 
complicates the task.

The Ising model \cite{Ising} is the fundamental model in the theory of 
critical phenomena. Its theoretical importance became evident in 1944, when
Onsager was able to compute the free energy on the square lattice in 
absence of magnetic field, providing in this way the first exact description 
of a second order phase transition \cite{Onsager}. It had to become clear later
that the Ising model corresponds to the simplest universality class of critical
behaviour. Since Onsager's work, the Ising model has been an essential
indicator of the progress in the analytic study of critical phenomena. While it
remains unsolved in three dimensions, the lattice studies gave additional 
exact results in the two dimensional case. In 
1952 Yang published the first derivation of the formula for the spontaneous 
magnetisation that Onsager had presented three years before \cite{Yang}. 
Scaling theory would later show that this result amounted to completing the 
list of critical exponents of the Ising universality class. 
Further important progress was made on the determination of correlation 
functions \cite{McCoyWu}. It was found, in particular, that the spin-spin 
correlator can be expressed, in the scaling limit, through the solution of a 
differential equation of Painlev\'e type \cite{WMcTB}. Up to some 
generalisations, this remains the only non-trivial correlation function of
quantum field theory to be exactly known.

These results refer to zero magnetic field, the Ising model in a field having 
never been solved on the lattice. For this reason, 
A.~Zamolodchikov's solution of the model with magnetic field, at the critical
temperature, {\it directly in the scaling limit}, came as a major surprise
for many \cite{Taniguchi}. Even more so since this solution consisted of a
long list of scattering amplitudes for eight different species of relativistic
particles. The most striking aspect of Zamolodchikov's work, however, was 
that, apart from its consequences for the Ising model, it actually implied new
exact results for all the universality classes of critical behaviour in two 
dimensions. In fact, if conformal field theory had given a complete description
of critical points \cite{BPZ}, now it was shown that some renormalisation group
trajectories flowing out of each critical point correspond to exactly solvable
(integrable) quantum field theories. These describe particular directions in
the scaling region of statistical models which, in general, are not solvable as
long as the non-universal lattice details are not eliminated through the 
scaling limit. Specific realisations of a given universality class, however,
can be solvable already on the lattice. In particular, a solvable lattice
model yielding the same scaling limit of the Ising model in a magnetic field
at critical temperature was found in \cite{WNS}.

Dealing directly with the scaling limit, integrable quantum field theory is 
the most effective tool for extracting exact information about universality
classes. In this context, `integrable' means that the 
relativistic scattering theory associated to the quantum field theory can be
determined exactly. Once this has been done, the next task is that of bridging
the gap between the scattering solution and the quantities of more direct 
interest for statistical mechanics. In this article we review the results
obtained through this approach for the universality class of the Ising model in
a magnetic field, on the infinite plane. Although we refrain from making 
reference to other models, we stress the general reasons which allow to obtain 
similar results for the other universality classes in two dimensions. Since we
always work in the continuum limit, we take care of comparing the field 
theoretical predictions for the universal quantities with the available lattice
estimates. 

The article is organised as follows. In the next section we recall the 
definition of the Ising model on the lattice before turning to the field 
theoretical description of the critical point and the scaling region around it.
In section~3 we review the
origin of integrable quantum field theories, their solution in the scattering
framework and the form factor approach to the computation of correlation 
functions. All this is illustrated in practice through the application to the 
integrable directions of the Ising field theory in section~4, where the known 
results for the purely thermal case and the recent advances for the magnetic 
case are presented within the same framework. In particular, we discuss the 
determination of form factors in the magnetic case \cite{immf,DS}. Section~5 
illustrates how the recent field theoretical
results reflect onto the traditional way of characterising critical behaviour 
and allow to complete the list of canonical amplitude ratios for the Ising 
universality class. Finally, in section~6, we briefly discuss how to exploit 
the integrable directions for a more general investigation of the scaling 
region and, in particular, review few basic results on the evolution of the
particle spectrum in the Ising field theory.

\resection{Ising field theory in two dimensions}
The Ising model is defined on the lattice by the reduced Hamiltonian
\EQ
E=-\frac{1}{T}\sum_{\langle i,j\rangle}\sigma_i\sigma_j-H\sum_i\sigma_i\,,
\label{lattice}
\EN
where $\sigma_i=\pm 1$ is the spin variable at the $i$-th site and the first 
sum is taken over nearest neighbours; the couplings $T$ and $H$ are 
referred to as temperature and magnetic field, respectively. The expectation
value of any lattice variable ${\cal O}$ is given by
\EQ
\langle{\cal O}\rangle=\frac1Z\sum_{\{\sigma_i\}}{\cal O}\,e^{-E}\,,
\EN
where
\EQ
Z=\sum_{\{\sigma_i\}}e^{-E}
\EN
is the partition function. We will always refer to the two-dimensional
ferromagnetic ($T>0$) case in the following.

Consider the case $H=0$. The Hamiltonian is then invariant under 
the change of sign of all spins. This spin reversal symmetry is broken 
spontaneously when $T$ is smaller than a critical value $T_c$ for which a 
second order phase transition takes place \cite{Onsager}. The critical point 
divides the temperature axis into a high-temperature, disordered phase,
and a low-temperature phase where a spontaneous magnetisation exists. 
The two phases are related by a duality transformation \cite{KWduality}.

No global symmetry is left in the model when $H\neq 0$, and that located at
$(T,H)=(T_c,0)$ in the $T$-$H$ plane is the only critical point in the model. 
Since changing the sign of $H$ simply amounts to a global spin reversal 
transformation, it is sufficient to refer to the case $H\geq 0$.

\vspace{.3cm}
The correlation length $\xi$ diverges at a second order phase transition
point and remains much larger than the lattice spacing in a neighbourhood 
of this point in coupling space. In the {\em scaling region} in which $T\goto
T_c$, $H\goto 0$, the system
can effectively be considered as translationally and rotationally invariant, 
and quantum field theory provides a continous description suitable for the 
investigation of the universal properties (see e.g. \cite{Cardybook}). 

In particular, the 
behaviour of the critical point on scales much larger than the lattice spacing
is described by a massless (mass $\sim 1/\xi$) field theory invariant under
scale transformations and, actually, under the larger group of conformal 
transformations. In two dimensions conformal symmetry is infinite dimensional
and, for this reason, conformal field theories are exactly solved \cite{BPZ}. 
As it should be, the Ising critical point is described by the simplest
(i.e. with the smallest operator content) conformal field theory satisfying 
the requirement of reflection positivity \cite{FQS}. This theory contains
three fundamental (primary) operators which are invariant (scalar) under
rotations, namely the identity $I$, the spin $\sigma(x)$ and the 
energy $\varepsilon(x)$ ($x=(x_1,x_2)$ denotes a point on the 
plane). The spin and energy operators are the continous version of the
lattice variables $\sigma_i$ and $\sum_j\sigma_i\sigma_j$ ($j$ nearest 
neighbour of $i$), respectively. Each primary operator possesses an infinite
number of `descendents', the simplest example being provided by the derivatives
of the primaries. A primary and its descendents share the same internal 
symmetry properties and form a `conformal family'. 

In a conformal field theory the product of two operators $\Phi_1$ and $\Phi_2$ 
(to be thought inside a correlation function) can be expanded over
a complete basis made of an infinite number of operators $A_k$ in the form
\EQ
\Phi_1(x)\Phi_2(0)=\sum_k C_{\Phi_1\Phi_2}^{A_k}
z^{\Delta_{A_k}-\Delta_{\Phi_1}-\Delta_{\Phi_2}}
\bar{z}^{\bar{\Delta}_{A_k}-\bar{\Delta}_{\Phi_1}-\bar{\Delta}_{\Phi_2}}
A_k(0)\,.
\label{ope}
\EN
Here $z=x_1+ix_2$ and $\bar{z}=x_1-ix_2$ are complex coordinates, the 
$C_{\Phi_1\Phi_2}^{A_k}$'s are called structure constants and $\Delta_\Phi$
and $\bar{\Delta}_\Phi$ are the conformal dimensions of an operator $\Phi(x)$.
The scaling dimension 
\EQ
X_\Phi=\Delta_\Phi+\bar{\Delta}_\Phi
\EN
and the euclidean spin
\EQ
s_\Phi=\Delta_\Phi-\bar{\Delta}_\Phi
\EN
determine the behaviour of the operator under scale transformations and 
rotations, respectively. The operators with a definite scaling dimension are 
called scaling operators. A scalar operator has $s_\Phi=0$.

While the scaling dimension of the identity operator vanishes, the
two non-trivial primary operators in Ising field theory have dimensions
$X_\sigma=1/8$ and $X_\varepsilon=1$. These two numbers determine all the 
critical exponents of the Ising universality class (see section~5). 
The conformal dimensions of the descendent operators differ by positive 
integers from those of the corresponding primary.

The structure of the operator product expansion in the Ising conformal theory
can be symbolically expressed as
\bea
&& \sigma\,\times\,\sigma\sim [I]+[\varepsilon]\,\nonumber\\
&& \sigma\,\times\,\varepsilon\sim[\sigma]\,\label{isingope}\\
&& \varepsilon\,\times\,\varepsilon\sim [I]\,,\nonumber
\eea
where the square brackets indicate the appearence of a whole conformal family
on the r.h.s. Clearly, this structure is compatible with the fact that
the conformal families $[I]$ and $[\varepsilon]$ are even under spin reversal
while $[\sigma]$ is odd. 

The Ising critical point is left invariant by the duality transformation 
which exchanges the high- and low-temperature phases at $H=0$. While the
energy operator $\varepsilon(x)$, which drives the model away from criticality 
along the thermal axis, changes sign under duality (this is why $[\varepsilon]$
does not appear in the last of (\ref{isingope})), such a transformation maps
$\sigma(x)$ onto a disorder operator $\mu(x)$ \cite{KC}. The operators
$\sigma$ and $\mu$ have the same scaling dimension $1/8$, but are mutually
non-local, in the following sense.
Two operators $\Phi_1$ and $\Phi_2$ are said to be mutually local if their 
product is unchanged when one of them is taken once around the other on the 
plane (i.e. under the analytic continuation $z\goto e^{2i\pi}z$, 
$\bar{z}\goto e^{-2i\pi}\bar{z}$ in (\ref{ope})). Of course this amounts to
a statement about the single-valuedness of correlation functions involving 
the two operators. An operator is said to be local if it is local with
respect to itself. It can be shown that the local operators (the only ones of 
interest for us) are those with integer or half-integer euclidean spin. 
The mildest type of mutual non-locality (called semi-locality) 
corresponds to the case
\EQ
\langle\cdots\Phi_1(e^{2i\pi}z,e^{-2i\pi}\bar{z})\Phi_2(0)\cdots\rangle=
l_{\Phi_1,\Phi_2}\langle\cdots\Phi_1(z,\bar{z})\Phi_2(0)\cdots\rangle\,,
\label{locality}
\EN
where $l_{\Phi_1,\Phi_2}$ is a phase called semi-locality factor.

The leading term in the $x\goto 0$ expansion of the product $\sigma\times\mu$
is \cite{KC}
\EQ
\sigma(x)\mu(0)\sim |x|^{-1/4}[\sqrt{z}\,\psi(0)+
\sqrt{\bar{z}}\,\bar{\psi}(0)]+\ldots\,,
\label{sigmamu}
\EN
where $\psi$ and $\bar{\psi}$ have conformal dimensions 
$(\Delta,\bar{\Delta})=(1/2,0)$ and $(0,1/2)$, respectively (the terms omitted
in the r.h.s. contain descendents of $\psi$ and $\bar{\psi}$). This means that
taking $\sigma$ around $\mu$ once produces a minus sign, so that the two
operators are semi-local with $l_{\sigma,\mu}=-1$. Similarly, $\psi$ and 
$\bar{\psi}$ are semi-local with respect to $\sigma$ and $\mu$ with the same
semi-locality factor $-1$.

\begin{figure}
\centerline{
\includegraphics[width=7cm]{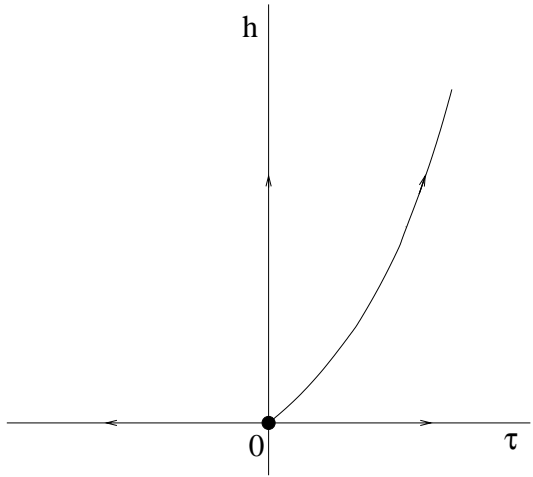}}
\caption{Coupling space of the Ising field theory (\ref{A}). The oriented lines
indicate some renormalisation group trajectories flowing out of the critical 
point at the origin. The integrable directions coincide with the principal 
axes. In our conventions $\tau>0$ corresponds to $T>T_c$.}
\end{figure}

Summarising, the three conformal families $[I]$, $[\sigma]$ and $[\varepsilon]$
we originally considered form a `local section' of the Ising conformal field
theory, namely a maximal set of fields all mutually local and closed under 
operator
product expansion. Duality leads to consider a second local section differing
from the first one by the substitution of $[\sigma]$ with $[\mu]$. A last 
local section is given by $[I]$, $[\psi]$, $[\bar{\psi}]$ and $[\varepsilon]$.
All the conformal families together form the full space of local operators
of the Ising conformal theory.

The operators $\psi$ and $\bar{\psi}$ are identified by their conformal
dimensions as the two components of a neutral (Majorana) fermion. This is 
why the Ising conformal point is described by the {\em free}\footnote{No 
interaction term preserving scale invariance can be formed.} Hamiltonian
(or euclidean action)
\EQ
{\cal A}_0=\frac12\int d^2x\,(\psi\bar{\partial}\psi+
\bar{\psi}\partial\bar{\psi})\,,
\label{A0}
\EN
where $\partial=\partial_z=(\partial_1-i\partial_2)/2$ and 
$\bar{\partial}=\partial_{\bar{z}}=(\partial_1+i\partial_2)/2$. The conformal
dimensions $(1/2,1/2)$ of the energy operator show that it is  
bilinear in the fermion components: 
\EQ
\varepsilon\sim\bar{\psi}\psi\,;
\label{energy}
\EN
$\sigma$ and $\mu$, on the contrary, are semi-local with respect to $\psi$ 
and $\bar{\psi}$ and give rise to the non-trivial sector of the theory
(\ref{A0}).

\vspace{.3cm}
The field theory describing the scaling region around the critical point
is obtained by adding to the conformal action (\ref{A0}) the contributions
of the operators conjugated to the termperature and the magnetic field,
namely the energy and the spin operator, respectively. This leads to the 
{\em Ising field theory}
\EQ
{\cal A}={\cal A}_{0}-\tau\int d^2x\,\var(x)-h\int d^2x\,\sigma(x)\,,
\label{A}
\EN
where
\bea
&& \tau\sim M^{2-X_\var}=M\,,\nonumber \\
&& h\sim M^{2-X_\sigma}=M^{15/8}\nonumber
\eea
are dimensional couplings measuring the deviation from critical temperature
and the magnetic field, respectively. Here, $M\sim 1/\xi$ denotes the mass 
scale associated to the breaking of scale invariance away from criticality.
It is worth stressing that the action (\ref{A}) is uniquely selected as the 
scaling limit of (\ref{lattice}) by the fact that $\sigma$ and $\varepsilon$ 
are the only (non-constant) scalar relevant\footnote{In the renormalisation 
group language, an operator is relevant if its scaling dimension is smaller 
than the space dimensionality (2 in our case).} operators in the local section 
of the operator space containing $\sigma$. Additional terms in (\ref{A}) play
a role only when trying to account for subleading terms in the expansion of 
lattice observables around the critical point (corrections to scaling). 
The field theory (\ref{A}) describes a one-parameter family of renormalisation 
group trajectories flowing out of the critical point at $\tau=h=0$ (Figure~1) 
and labelled by the dimensionless quantity 
\EQ
\eta=\frac{\tau}{|h|^{8/15}}\,\,.
\label{eta}
\EN

The Ising model with $H=0$ is solvable on the lattice and then must be solvable
in the scaling limit. In fact, (\ref{A0}) and (\ref{energy}) imply that for 
$h=0$ the action (\ref{A}) describes a {\em free massive} fermion, the mass 
being proportional to $|\tau|$.
For $H\neq 0$ the Ising model has never been solved on the lattice. 
A.~Zamolodchikov showed that it is solvable directly in the scaling limit
if $T=T_c$ \cite{Taniguchi}. This is a consequence of the fact that (\ref{A}) 
with $\tau=0$ is an integrable quantum field theory. We summarise in the
next section some generalities about integrable field theories before turning
to the study of the integrable directions in the scaling Ising model (of course
the free theory resulting from (\ref{A}) when $h=0$ is a particularly simple 
case of integrability).

\resection{Integrable quantum field theories}
\subsection{Conserved currents}
The notion of integrability is generally associated to the presence
of an infinite number of conserved quantities. In two-dimensional quantum
field theory a conservation law takes the form
\EQ
\bar{\partial}T_{s+1}=\partial\Theta_{s-1}\,,
\label{conservation}
\EN
where $T_{s+1}$ and $\Theta_{s-1}$ are local operators (currents) with spin
$s+1$ and $s-1$, respectively. Any quantum field theory possesses the 
conservation law (\ref{conservation}) with $s=1$, $T_{2}$ and $\Theta_{0}$ 
being components of the energy-momentum tensor. 
Are there theories allowing for additional, non-trivial conservation laws? 
The answer is obviously affirmative for the conformal theories: any 
descendent $T_s$ of the identity with conformal dimensions $(s,0)$ is a local 
operator satisfying (\ref{conservation}) with zero on the r.h.s. Of course,
this is a direct consequence of the infinite dimensional character of conformal
symmetry in two dimensions. It is then natural to ask whether any conservation 
law (other than that of energy and momentum) can  
survive when a non-scale-invariant theory is obtained as a perturbation of a 
conformal action ${\cal A}_{CFT}$ by a relevant operator $\Phi$, namely when 
considering the action
\EQ
{\cal A}={\cal A}_{CFT}-g\int d^2x\,\Phi(x)\,\,.
\label{perturbed}
\EN
In this case the original (conformal) conservation laws get modified into 
\cite{Taniguchi}
\EQ
\bar{\partial}T_{s}=gR^{(1)}_{s-1}+\cdots+g^nR^{(n)}_{s-1}+\cdots\,,
\label{perturbative}
\EN
where $R^{(n)}_{s-1}$ are operators with conformal dimensions 
$(s-n(1-\Delta),1-n(1-\Delta))$ (the perturbing operator $\Phi$ has 
dimensions $(\Delta,\Delta)$, $\Delta<1$). It follows that in any theory with
a spectrum of conformal dimensions bounded from below the r.h.s. of 
(\ref{perturbative}) can only contain a finite number of terms. Moreover,
the operators 
$R^{(n)}_{s-1}$ with $n>1$ can be accomodated within the operator space of 
a theory with a discrete spectrum of conformal dimensions only if special 
relations between the dimensions are fulfilled. Hence, in a generic
case, the r.h.s. of (\ref{perturbative}) contains only the operator 
$R^{(1)}_{s-1}$, which is a descendent of the perturbing operator $\Phi$. 
Thus the issue of conservation away from criticality is reduced to 
establishing under which conditions (if any) the operator $gR^{(1)}_{s-1}$ can 
be written in a total derivative form $\partial\Theta_{s-2}$. The
complete characterisation of the operator space provided by conformal
field theory\footnote{It is generally assumed that the operator spaces
at the conformal point and in the perturbed theory (\ref{perturbed}) are 
isomorphic.} allows the identification of sufficient conditions through a 
so-called `counting argument' \cite{Taniguchi}. This exploits notions of 
conformal field theory which are not required in the remainder of this article,
and we prefer to directly state the result of the analysis, which is 
remarkable: several non-trivial conservation laws (expected to be the first 
representatives of infinite series) can be found for a number of different 
perturbations of essentially all the known conformal points in two-dimensions;
each integrable direction in coupling space is characterised by a specific set 
of values of the spin $s$ for which a conservation law of the form 
(\ref{conservation}) is present.

For the Ising field theory (\ref{A}), in particular, the counting argument 
implies integrability when $\tau=0$ and, of course, when $h=0$. The theory is 
not integrable when both couplings are different from zero \cite{Giuseppe,
Fonseca}.

\subsection{Scattering theory}
The field theory obtained by perturbing a conformal theory by one or more 
relevant operators normally develops a finite correlation length and admits
a description in terms of massive particles. These particles propagate in a
space with one spatial and one time dimension, related by analytic continuation
to imaginary time to the euclidean plane we have in mind for the applications
to equilibrium statistical mechanics. 

Integrability induces major simplifications in the scattering of relativistic
particles (see \cite{ZZ} and references therein). Call
\EQ
P_s=\int_{-\infty}^{+\infty}dx_1\,[T_{s+1}(x)+\Theta_{s-1}(x)]
\label{Ps}
\EN
the spin $s>0$ conserved quantities, $P_1$ being the sum of energy and 
momentum. Also, denote by $A_a(p^\mu)$ a particle of type $a$ and 
energy-momentum $p^\mu=(p^0,p^1)$ satisfying the mass shell condition 
\EQ
p^\mu p_{\mu}=(p^0)^2-(p^1)^2=m_a^2\,\,.
\label{shell}
\EN
A conserved quantity acts as
\EQ
P_s|A_{a_1}(p_1^\mu)\ldots A_{a_n}(p_n^\mu)\rangle_{in\,\,(out)}=
\left(\sum_{k=1}^n\omega_s^{a_k}(p^\mu_k)\right)
|A_{a_1}(p_1^\mu)\ldots A_{a_n}(p_n^\mu)\rangle_{in\,\,(out)}
\EN
on the initial (final) state of a scattering process containing $n$ widely
separated particles. The behaviour under euclidean rotations fixes the form
of the one-particle eigenvalue to be
\EQ
\omega_s^{a}(p^\mu)=\kappa_s^ap^s\,,
\label{omega}
\EN
with $p=p^0+p^1$ and $\kappa_s^a$ a constant. Then conservation means that
\EQ
\sum_{k=1}^n\omega_s^{a_k}(p^\mu_k)=\sum_{j=1}^m\omega_s^{b_j}(q^\mu_j)
\EN
in a scattering process with $n$ particles $A_{a_k}(p^\mu_k)$ in the initial
state and $m$ particles $A_{b_j}(q^\mu_j)$ in the final state. Since this is
a system of an infinite number of equations (one for each conserved quantity
$P_s$) for a finite number of unknowns (the energies and momenta of the 
outgoing particles), one concludes that in a scattering process of an 
integrable quantum field theory 

\noindent
i) the final set of energies and momenta coincides with the initial one. 

\begin{figure}
\centerline{
\includegraphics[width=8cm]{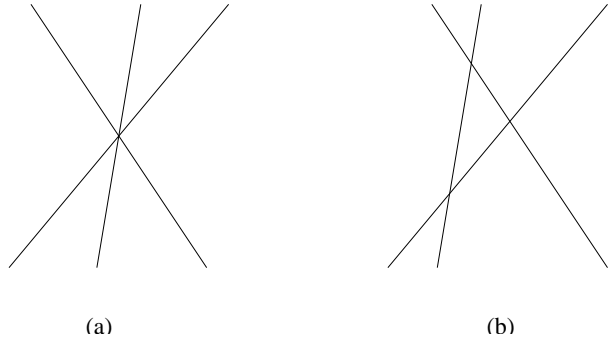}}
\caption{Space-time diagrams for a three-particle scattering process. A genuine
three-body collision (a) or a sequence of widely separated two-body collisions 
(b) can be obtained by a suitable choice of initial conditions on the wave 
packets.}
\end{figure}

This is a remarkable simplification of the relativistic scattering problem. 
There is, however, an additional important result whose origin can be 
understood through the following euristic argument\footnote{See 
\cite{Iagolnitzer} for a discussion in the axiomatic framework.} \cite{SW}. 
The conserved quantities $P_s$ can be seen
as generators of space-time displacements on wave packets. While $P_1$ simply
translates the trajectories of all particles by the same amount in space-time,
for $s>1$ the amount of the shift depends on the momentum of each particle. 
Consider now a state containing three particles (wave packets) with different 
momenta and initial conditions chosen in such a way that they give rise to a 
genuine three-body interaction (Figure 2a). We can perform on this state a 
suitably chosen transformation generated by the $P_s$ with $s>1$ which shifts 
each trajectory by a different amount and resolves the three-body interaction 
into a sequence of three two-body collisions widely separated in space-time 
(Figure 2b). Since the $P_s$ are conserved, the transformation commutes with 
the time evolution and then the scattering amplitudes for the two processes 
coincide. The general conclusion is that in integrable quantum field theory

\noindent
ii) any $n$-particle scattering amplitude factorises into the product of
$n(n-1)/2$ two-particle amplitudes.

We discussed the elasticity (i.e. absence of particle production) and 
factorisation of the scattering assuming the 
presence of an infinite number of conserved quantities. It has been shown,
however, that the existence of one such a quantity besides energy-momentum
is sufficient to arrive at the same conclusions \cite{Parke}. 

\begin{figure}
\centerline{
\includegraphics[width=4cm]{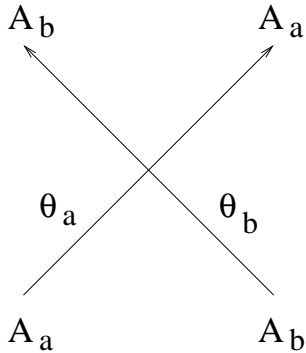}}
\caption{Space-time diagram associated to the scattering amplitude 
$S_{ab}(\theta_1-\theta_2)$; time runs upwards.}
\end{figure}

As a consequence of factorisation, the determination of the $S$-matrix (i.e. 
the collection of all scattering amplitudes) in integrable quantum field 
theories reduces to that of the two-particle amplitudes. For the sake of 
simplicity we restrict our discussion to the case of neutral particles,
which is enough for dealing with the Ising model. Then the two-particle 
amplitudes can be defined through the relation
\EQ
|A_a(\theta_1)A_b(\theta_2)\rangle=S_{ab}(\theta_1-\theta_2)
|A_b(\theta_2)A_a(\theta_1)\rangle\,,
\label{fz}
\EN
where we introduced the rapidity variables parameterising on-shell momenta
as 
\EQ
(p^0,p^1)=(m_a\cosh\theta,m_a\sinh\theta)
\label{onshell}
\EN
for a particle of mass $m_a$. Particles are ordered with rapidities 
decreasing (increasing) from left to right in initial (final) states . The 
dependence of the amplitudes on rapidity differences is a consequence of 
Lorentz invariance. A pictorial representation of a two-particle scattering 
amplitude is shown in Figure~3. Double application of 
(\ref{fz}) together with $S_{ab}(\theta)=S_{ba}(\theta)$ yield the unitarity 
relation 
\EQ
S_{ab}(\theta)S_{ab}(-\theta)=1\,\,.
\label{unitarity}
\EN
The crossing symmetry relation 
\EQ
S_{ab}(\theta)=S_{ab}(i\pi-\theta)
\label{crossing}
\EN
is a general property of relativistic scattering \cite{ELOP}.

The scattering amplitudes $S_{ab}(\theta)$ are meromorphic functions of the 
rapidity difference. A simple pole with residue
\EQ
S_{ab}(\theta\simeq iu_{ab}^c)\simeq\frac{i(\Gamma_{ab}^c)^2}{\theta-iu_{ab}^c}
\label{pole}
\EN
and $u_{ab}^c\in(0,\pi)$ corresponds to a bound state with mass square
\EQ
m_c^2=m_a^2+m_b^2+2m_am_b\cos u_{ab}^c
\label{triangle}
\EN
propagating in the $A_aA_b$ scattering channel. A three-particle coupling 
$\Gamma_{ab}^c$ is associated to each vertex of the corresponding diagram in 
Figure~4. It follows from (\ref{triangle}) that 
\EQ
\bar{u}_{ab}^c=\pi-u_{ab}^c
\EN
is the angle opposite to $m_c$ in a triangle with sides of length $m_a$, 
$m_b$, $m_c$. This implies the relation
\EQ
u_{ab}^c+u_{bc}^a+u_{ca}^b=2\pi\,\,.
\label{angles}
\EN
As a consequence of (\ref{crossing}), a crossed channel pole with negative 
residue at $\theta=i(\pi-u_{ab}^c)$ goes along with each pole (\ref{pole}).

\begin{figure}
\centerline{
\includegraphics[width=3cm]{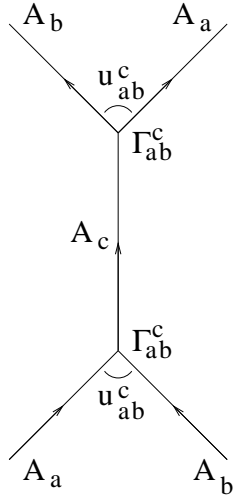}}
\caption{Simple pole diagram associated to equation (\ref{pole}).}
\end{figure}

The general meromorphic solutions of (\ref{unitarity}) and (\ref{crossing}) 
are of the form \cite{Mitra}
\EQ
S_{ab}(\theta)=\pm\prod_{\alpha\in{\cal A}_{ab}}t_\alpha(\theta)\,,
\label{Sab}
\EN
with 
\EQ
t_\alpha(\theta)=\frac{\tanh\frac{1}{2}(\theta+i\pi\alpha)}
                        {\tanh\frac{1}{2}(\theta-i\pi\alpha)}\,\,.
\label{talpha}
\EN
Real values of $\alpha$ between $0$ and $1$ correspond to the bound state poles
(\ref{pole}). Important constraints on the bound state structure come from 
the fact that an infinite number of quantities besides energy and momentum
has to be conserved at each three-particle vertex \cite{Taniguchi}. In the 
vicinity of a bound state pole we can write
\EQ
|A_a(\theta+i\bar{u}_{ca}^b-\epsilon)A_b(\theta-i\bar{u}_{bc}^a+\epsilon)
\rangle\sim\frac1\epsilon\,|A_c(\theta)\rangle\,\,.
\EN
Applying $P_s$ to both sides and equating the eigenvalues gives
\EQ
\kappa_s^am_a^se^{is\bar{u}_{ca}^b}+\kappa_s^bm_b^se^{-is\bar{u}_{bc}^a}=
\kappa_s^cm_c^s\,\,.
\label{abc}
\EN

Integrability in presence of bound states also provides functional relations
between different amplitudes. Indeed, we can consider the scattering of a 
particle $A_d$ with a resonant pair $A_aA_b$ and exploit the freedom of 
translating the particle trajectories to obtain (Figure~5)
\EQ
S_{dc}(\theta)=S_{da}(\theta-i\bar{u}_{ac}^b)S_{db}(\theta+i\bar{u}_{bc}^a)
\,\,.
\label{bootstrap}
\EN
These equations are known as `bootstrap' equations for the following reason.
Suppose that the use of (\ref{abc}) plus other considerations ends up in an
educated guess for the scattering amplitudes of the `elementary' (lightest) 
particles of the theory. Then one can use (\ref{bootstrap}) with $A_a$, $A_b$, 
$A_d$ `elementary' particles and $A_c$ one of their bound states to determine
the amplitude $S_{dc}$. Each amplitude computed in this way may present poles 
corresponding to new particles, in which case the use of (\ref{bootstrap}) is 
iterated. Once this bootstrap procedure consistently reaches a point where no 
new particles are generated, one is left with the exact solution of an 
integrable quantum field theory (see \cite{Giuseppe} for a series of 
examples).

\begin{figure}
\centerline{
\includegraphics[width=10cm]{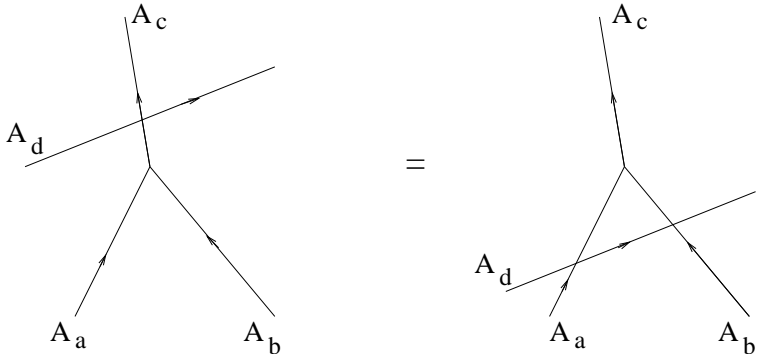}}
\caption{Pictorial representation of the bootstrap equation (\ref{bootstrap}).}
\end{figure}

\subsection{Correlation functions}
Having the $S$-matrix is not enough for statistical mechanical applications;
correlation functions are needed. These can be expressed in the form of 
spectral sums over complete sets of particle states. For example, the 
(euclidean) correlator of two scalar operators reads\footnote{Particle states
are normalised through the condition $\langle A_a(\theta_1)|A_b(\theta_2)
\rangle=2\pi\delta_{ab}\delta(\theta_1-\theta_2)$.}
\EQ
\langle\Phi_1(x)\Phi_2(0)\rangle=
\sum_{n=0}^\infty\frac{1}{(2\pi)^n}\int_{\theta_1>\cdots>\theta_n}
d\theta_1\cdots d\theta_n\,
F^{\Phi_1}_{a_1\ldots a_n}(\theta_1,\ldots,\theta_n)
\left[F^{\Phi_2}_{a_1\ldots a_n}(\theta_1,\ldots,
\theta_n)\right]^*e^{-|x|E_n}\,,
\label{spectral}
\EN
where
\EQ
E_n=\sum_{k=1}^nm_{a_k}\cosh\theta_k\,,
\EN
and the matrix elements (Figure~6)
\EQ
F^\Phi_{a_1\ldots a_n}(\theta_1,\ldots,\theta_n)=\langle0|\Phi(0)|
A_{a_1}(\theta_1)\ldots A_{a_n}(\theta_n)\rangle
\label{formfactors}
\EN
are called {\em form factors} ($|0\rangle$ is the vacuum state).

The form factors can be computed exactly in integrable quantum field 
theory. They are subject to a number of equations \cite{KW,Smirnov} which for 
the case of neutral particles we are discussing read
\bea
&& F^\Phi_{a_1\ldots a_n}(\theta_1+\Lambda,\ldots,\theta_n+\Lambda)=
e^{s\Lambda}F^\Phi_{a_1\ldots a_n}(\theta_1,\ldots,\theta_n)
\label{ff0}\\
&& F^\Phi_{a_1\ldots a_ia_{i+1}\ldots a_n}(\theta_1,\ldots,\theta_i,
\theta_{i+1},\ldots,\theta_n)=\nonumber\\
&& \hspace{3.5cm}S_{a_ia_{i+1}}(\theta_i-\theta_{i+1})
F^\Phi_{a_1\ldots a_{i+1}a_{i}\ldots a_n}(\theta_1,\ldots,\theta_{i+1},
\theta_{i},\ldots,\theta_n)
\label{ff1}\\
&& \mbox{Res}_{\theta_a-\theta_b=iu_{ab}^c}\,
F^\Phi_{aba_1\ldots a_n}(\theta_a,\theta_b,\theta_1,\ldots,\theta_n)=
i\Gamma_{ab}^c\,F^\Phi_{ca_1\ldots a_n}(\theta_c,\theta_1,\ldots,\theta_n)
\label{ff2}\\
&& F^\Phi_{a_1\ldots a_n}(\theta_1+2i\pi,\theta_2,\ldots,\theta_n)=
l_{\Phi,\phi_{a_1}}F^\Phi_{a_2\ldots a_na_1}(\theta_2,\ldots,\theta_n,\theta_1)
\label{ff3}\\
&& \mbox{Res}_{\theta'=\theta+i\pi}\,
F^\Phi_{aba_1\ldots a_n}(\theta',\theta,\theta_1,\ldots,\theta_n)=
\nonumber\\
&& \hspace{4cm}i\delta_{ab}\left(1-l_{\Phi,\phi_a}
\prod_{j=1}^nS_{a_ja}(\theta_j-\theta)
\right)F^\Phi_{a_1\ldots a_n}(\theta_1,\ldots,\theta_n)\,\,.
\label{ff4}
\eea

\begin{figure}
\centerline{
\includegraphics[width=4cm]{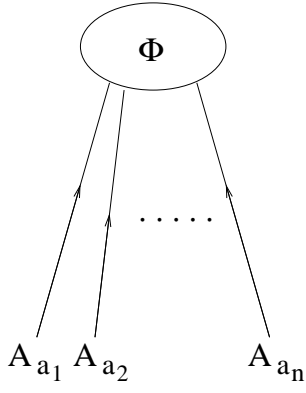}}
\caption{Pictorial representation of an $n$-particle form factor.}
\end{figure}

The first equation expresses Lorentz covariance for a spin $s$ operator, 
while the second is a consequence of (\ref{fz}) and factorisation of 
multi-particle scattering. The third equation says that the form factors
inherit the direct channel bound state poles of the $S$-matrix 
($\theta_c=(\bar{u}_{bc}^a\theta_a+\bar{u}_{ca}^b\theta_b)/u_{ab}^c$).

Within the framework of relativistic scattering theory, a crossing process
corresponds to trading a particle of energy-momentum $p^\mu$ in the initial 
(final) state with an antiparticle of energy-momentum $-p^\mu$ in the final 
(initial) state \cite{ELOP}. The inversion of energy-momentum corresponds to 
an $i\pi$ shift in the rapitidy parameterisation. Hence, equation (\ref{ff3})
shows that the (kinematically immaterial) double crossing of a particle 
corresponds to a reordering of rapidities. If the operator $\phi_a$ 
which creates the particle\footnote{Any operator with $\langle 0|\phi_a|A_a
\rangle\neq 0$ can be taken as creating operator. In order to get rid of spin
factors, we always choose $\phi_a$ to be scalar.} $A_a$ is non-local 
(semi-local, for simplicity) with respect to $\Phi$, the double crossing of 
this particle also produces a phase factor $\l_{\Phi,\phi_a}\neq 1$ (recall 
the definition (\ref{locality})) \cite{YZ}. 

Finally, equation (\ref{ff4}) expresses the fact that two identical particles
with opposite energy-momentum can annihilate. This gives rise
to a pole whose residue reflets the fact that the annihilation of adjacent
particles can take place either directly or through the analytic continuation
(\ref{ff3}).

The form factor equations listed above are satisfied by all the operators of 
the theory and then admit an infinite number of solutions. Identifying the 
solution corresponding to a given operator is a central problem in this
approach. The following argument turns out to be quite helpful in this respect.
Consider a scalar operator $\Phi$ with scaling dimension $X_\Phi$ in a massive
theory. Since
\EQ
\langle\Phi(x)\Phi(0)\rangle\sim\frac{1}{|x|^{2X_\Phi}}\,,\hspace{1cm}
|x|\goto 0
\EN
we know that\footnote{We denote by $\langle\cdots\rangle_c$ the connected 
correlators.}
\EQ
M_p=\int d^2x|x|^p\langle\Phi(x)\Phi(0)\rangle_c
\label{Mp}
\EN
is finite only if 
\EQ
p+2>2X_\Phi\,\,.
\label{bound1}
\EN
Once we use (\ref{spectral}) to expand the correlator in (\ref{Mp}) and perform
the spatial intergration, we are left with a series of $n$-fold integrals over 
rapidities with positive integrands (moduli squared of form factors). Provided
(\ref{bound1}) holds, each term of the series has to be finite, what implies
an upper bound on the asymptotic behaviour of the form factors. One can 
check that chosing the value of $p$ yielding the most stringent bound gives
the final result \cite{immf}
\EQ
\lim_{|\theta_i|\rightarrow\infty}F^{\Phi}_{a_1\ldots a_n}
(\theta_1,\ldots,\theta_n)\sim\exp(Y_\Phi|\theta_i|)
\label{asymptotics}
\EN
with
\EQ
Y_\Phi\leq\frac{X_\Phi}{2}\,\,.
\label{bound}
\EN

Explicit expressions for all the form factors of a given operator can be 
obtained in many cases. The problem of summing series like (\ref{spectral}), 
however, remains unsolved for interacting theories. Due to the exponential
dumping factor, (\ref{spectral}) is a large distance expansion. A complementary
short distance expansion is provided by perturbation theory around the 
conformal point \cite{AlyoshaYL}. Direct comparison shows that both expansions 
converge very rapidly and that the first few terms provide a very good 
matching at the intermediate scales where the crossover from the short
distance power law behaviour to the large distance exponetial decay takes 
place. 

Integrated correlators like (\ref{Mp}) are what one needs to compute in many 
physical applications (see e.g. section~5). In these cases the contribution 
of short distances is suppressed by the powers of $|x|$ in the integrand, and 
the form factor 
expansion alone is sufficient to obtain precise estimates. A check of the 
convergence can be done through sum rules allowing to recover conformal data
from the off-critical correlation functions. Conformal theories in two
dimensions are labelled by a number $C$ called `central charge' \cite{BPZ}. It 
is a consequence of Zamolodchikov's $C$-theorem \cite{Zamocth} that the 
central charge of the conformal limit of a massive theory can
be expressed as \cite{Cardycth}
\EQ
C=\frac{3}{4\pi}\int d^2x\,|x|^2\langle
\Theta(x)\Theta(0)\rangle_{c}\,,
\label{cth}
\EN
where $\Theta(x)$ is the trace of the energy-momentum tensor. Similarly, the 
scaling dimension of a relevant operator $\Phi(x)$ is computable through the 
sum rule \cite{DSC}
\EQ
X_\Phi=-\frac{1}{2\pi\langle\Phi\rangle}
\int d^2x\langle\Theta(x){\Phi}(0)\rangle_{c}\,\,.
\label{Xth}
\EN

\resection{Integrable directions in the scaling Ising model}

\subsection{Zero magnetic field}
As we have already seen, the scaling limit of the Ising model without 
magnetic field is described by the action (\ref{A}) with $h=0$, namely by
a theory of free neutral fermions with mass $m$ proportional to $|\tau|$. 
Depending on the sign of $\tau$, this theory describes the ordered or 
disordered phase. To be definite, we will work in the high-temperature phase 
($\tau>0$) and will use the duality transformations $\varepsilon\leftrightarrow
-\varepsilon$ and $\sigma\leftrightarrow\mu$ to obtain the results in the 
ordered phase.

The free fermionic theory provides a particularly simple example of integrable
field theory. It contains a single neutral particle $A$ with fermionic 
statistics and $S$-matrix equal to $1$. In order to fit the conventions on form
factors of the previous section, however, we will think of the particle as 
created by a bosonic operator and absorb the anticommutativity into the 
$S$-matrix, so that the two-particle scattering amplitude reads
\EQ
S=-1\,\,.
\label{fermions}
\EN
In the high-temperature phase we are considering, the particles correspond to
local spin excitations and the spin operator $\sigma$ is naturally identified
as the scalar creation operator\footnote{In the low-temperature phase the 
excitations are kinks interpolating between the two degenerate vacua. Such
topologic excitations are non-local in the spin degrees of freedom and the
creation operator corresponding to the amplitude (\ref{fermions}) is the 
disorder operator $\mu$.}. 

For the present case in which we have only one type of particle the notation
for form factors can be simplified to
\EQ
F^\Phi_{n}(\theta_1,\ldots,\theta_n)=\langle0|\Phi(0)|
A(\theta_1)\ldots A(\theta_n)\rangle\,,
\EN
so that the form factor equations become
\bea
&& F^\Phi_{n}(\theta_1+\Lambda,\ldots,\theta_n+\Lambda)=
e^{s\Lambda}F^\Phi_{n}(\theta_1,\ldots,\theta_n)
\label{fn0}\\
&& F^\Phi_{n}(\theta_1,\ldots,\theta_i,
\theta_{i+1},\ldots,\theta_n)=-F^\Phi_{n}(\theta_1,\ldots,\theta_{i+1},
\theta_{i},\ldots,\theta_n)
\label{fn1}\\
&& F^\Phi_{n}(\theta_1+2i\pi,\theta_2,\ldots,\theta_n)=
l_{\Phi,\sigma}F^\Phi_{n}(\theta_2,\ldots,\theta_n,\theta_1)
\label{fn3}\\
&& \mbox{Res}_{\theta'=\theta+i\pi}\,
F^\Phi_{n+2}(\theta',\theta,\theta_1,\ldots,\theta_n)=i\left[1-(-1)^n
l_{\Phi,\sigma}\right]F^\Phi_{n}(\theta_1,\ldots,\theta_n)
\label{fn4}
\eea
(equation (\ref{ff2}) plays no role in absence of bound states). We know from
the discussion of section~2 that 
\EQ
l_{\sigma,\sigma}=l_{\varepsilon,\sigma}=-l_{\mu,\sigma}=1\,\,.
\EN
The solutions \cite{BKW}
\bea
&& F^\Theta_n(\theta_1,\ldots,\theta_n)=
c_1m\,F^\varepsilon_n(\theta_1,\ldots,\theta_n)=
-2i\pi\,m^2\delta_{n,2}\,\sinh\frac{\theta_1-\theta_2}{2}
\label{theta2}\\
&& F^\sigma_{2n+1}(\theta_1,\ldots,\theta_{2n+1})=i^n F^\sigma_1
\prod_{i<j}\tanh\frac{\theta_i-\theta_j}{2}
\label{sigmaodd}\\
&& F^\mu_{2n}(\theta_1,\ldots,\theta_{2n})=i^nF^\mu_0\prod_{i<j}
\tanh\frac{\theta_i-\theta_j}{2}
\label{mueven}
\eea
are the only ones satisfying the asymptotic bound (\ref{bound}) for the case of
relevant scalar operators. This is a simple example of how the form factor 
approach reveals the operator content behind a scattering theory. A counting of
solutions including descendents confirms the correspondence with the operator
space of the conformal point \cite{CM}. In writing (\ref{theta2}) we
used the fact that, being the operator which breaks conformal invariance, 
$\varepsilon$ is proportional to the trace of the energy-momentum tensor
($c_1$ is a dimensionless constant) for which the normalisation condition
\EQ
F^\Theta_{aa}(\theta+i\pi,\theta)=2\pi m_a^2
\label{thetanorm}
\EN
holds. The vanishing of the $F^\varepsilon_n$ with $n\neq 2$ 
matches the fact that the energy operator is bilinear in the free fermions. 
The results for $\sigma$ and $\mu$ are non-trivial and reflect the non-locality
of these operators with respect to the fermions. The form factors of 
$\sigma$ with an even number of particles have to vanish because the particles 
are odd under $\sigma\goto -\sigma$; the invariance of $\mu$ under this 
symmetry induces the vanishing of the $F^\mu_{2n+1}$.

It follows from (\ref{theta2}) that the form factor expansion of a two-point 
function involving the energy operator contains only one term. In particular, 
the sum rules (\ref{cth}) and (\ref{Xth}) immediately give the exact 
results\footnote{The integral in (\ref{Xth}) diverges for $\Phi=\varepsilon$.
See \cite{DSC} on this point.} ($F^\mu_0=\langle\mu\rangle$)
\EQ
C=\frac32\int_0^\infty d\theta\,\frac{\sinh^2\theta}{\cosh^4\theta}=\frac12\\
\EN
\EQ
X_{\mu}=\frac{1}{2\pi}\int_0^\infty d\theta\,\frac{\sinh^2\theta}
{\cosh^3\theta}=\frac18\,\,.
\EN

Form factor series with an infinite number of terms are instead obtained for 
the correlators $\langle\sigma\sigma\rangle$ and $\langle\mu\mu\rangle$. 
Due to the particularly simple form of the matrix elements (\ref{sigmaodd}) and
(\ref{mueven}) these series can be resummed \cite{BB,BL}. We quote here the 
final result of this procedure which, of course, reproduces that originally 
obtained in \cite{WMcTB} from the lattice solution. The correlators can be 
written in the form
\EQ
\langle\sigma(x)\sigma(0)\rangle=\langle\mu\rangle^2\,
e^{-\Sigma(t)}\sinh\frac12\chi(t/2)
\label{sigmasigma}
\EN
\EQ
\langle\mu(x)\mu(0)\rangle=\langle\mu\rangle^2\,
e^{-\Sigma(t)}\cosh\frac12\chi(t/2)\,,
\label{mumu}
\EN
where $t=m|x|$ and
\EQ
\Sigma(t)=\frac14\int_{t/2}^\infty d\rho\rho\left[(\partial_\rho\chi)^2-
4\sinh^2\chi\right]\,.
\EN
The function $\chi$ is the solution of the differential 
equation\footnote{Equation (\ref{painleve}) becomes a Painlev\'e III equation 
for the function $\eta=e^{-\chi}$.} 
\EQ
\partial^2_\rho\chi+\frac{1}{\rho}\partial_\rho\chi=2\sinh2\chi
\label{painleve}
\EN
satisfying the asymptotic conditions
\bea
&& \chi(\rho)\simeq -\ln\rho+constant\,,\hspace{1cm}\rho\goto 0\\
&& \chi(\rho)\simeq\frac2\pi K_0(2\rho)\,,\hspace{1cm}\rho\goto\infty\,\,.
\eea
Both (\ref{sigmasigma}) and (\ref{mumu}) behave as
\EQ
\frac{C_I}{|x|^{2X_\sigma}}=
\frac{C_I}{|x|^{1/4}}
\EN 
as $|x|\goto 0$. When $|x|\goto\infty$ the correlators decay exponentially: 
$\langle\sigma\sigma\rangle$ vanish while $\langle\mu\mu\rangle$ 
approaches $\langle\mu\rangle^2$. If the operators are normalised in such a
way that $C_I=1$, then
\EQ
\langle\mu\rangle=\pm m^{1/8}2^{1/12}e^{-1/8}A^{3/2}\,,
\label{spontaneous}
\EN
$A$ being the Glaisher constant
\EQ
A=2^{7/36}\pi^{-1/6}\exp\left[\frac13+\frac23\int_0^\frac12 dx\ln\Gamma(1+x)
\right]=1.282427129..\,\,.
\EN

These results for the correlation functions of the scaling Ising model without 
magnetic field were also derived in \cite{SMJ} using the theory of monodromy 
preserving deformations of ordinary differential equations. Very recently a 
simpler derivation based on the Ward identities associated to the conservation 
laws of the free fermionic theory has been given in \cite{FZ2}. The convergence
of the form factor series was analysed numerically in \cite{YZ}. A study of 
four-point functions in the form factor approach can be found in \cite{BNNPSW}.

\subsection{Non-zero magnetic field at $T=T_c$}
As anticipated in section~2, the Ising field theory (\ref{A}) with $h\neq 0$ 
is integrable only when $\tau=0$. The counting argument applied to this purely 
magnetic perturbation of the Ising conformal point shows that conserved 
quantities of the form (\ref{Ps}) exist for spin $s=1,7,11,13,17,19$ 
\cite{Taniguchi}. These are expected to be the first representatives of the 
infinite set 
\EQ
s=1,7,11,13,17,19,23,29\,\,\,(\mbox{mod}\,\,30)\,\,.
\label{e8}
\EN
It can be noted that 
this set of numbers coincides with the exponents of the Lie algebra $E_8$
repeated modulo the Coxeter number of the algebra. The spectrum of conserved
spins (\ref{e8}) and its relation with the algebra $E_8$ were first predicted 
by V.~Fateev \cite{FateevE8,FZE8}.

Since the Ising field theory with non-zero magnetic field does not possess any
internal symmetry, its mass spectrum is guaranteed to be non-degenerate. Hence
we know that the two-particle scattering amplidutes of the integrable theory 
must be of the form (\ref{Sab}). A.~Zamolodchikov looked for the minimal
solution\footnote{In general, the solution of the $S$-matrix equations of 
section~3.2 is not unique. There exists, however, a minimal solution 
possessing the smallest number of zeros and poles in the physical strip
Im$\theta\in(0,\pi)$.} of the bootstrap equations (\ref{bootstrap}) satisfying 
the conservation equations (\ref{abc}) with the set of spin values (\ref{e8})
\cite{Taniguchi}. He found that this requirement leads to a bootstrap system
closing on 8 particles $A_a$ ($a=1,\ldots,8$) with masses
\bea
m_2 &=& 2 m_1 \cos\frac{\pi}{5} = (1.6180339887..) \,m_1\nonumber\\
m_3 &=& 2 m_1 \cos\frac{\pi}{30} = (1.9890437907..) \,m_1\nonumber\\
m_4 &=& 2 m_2 \cos\frac{7\pi}{30} = (2.4048671724..) \,m_1\nonumber \\
m_5 &=& 2 m_2 \cos\frac{2\pi}{15} = (2.9562952015..) \,m_1\nonumber\\
m_6 &=& 2 m_2 \cos\frac{\pi}{30} = (3.2183404585..) \,m_1\nonumber\\
m_7 &=& 4 m_2 \cos\frac{\pi}{5}\cos\frac{7\pi}{30} = (3.8911568233..) \,m_1\
\nonumber\\
m_8 &=& 4 m_2 \cos\frac{\pi}{5}\cos\frac{2\pi}{15} = (4.7833861168..) \,m_1
\nonumber
\eea
in units of the lightest mass $m_1$. The scattering amplitudes can be written 
as
\EQ
S_{ab}(\theta)=\prod_{\gamma\in{\cal G}_{ab}}\left(t_{\gamma/30}(\theta)
\right)^{p_\gamma}
\EN
in terms of the building blocks (\ref{talpha});
the complete list of indices $\gamma$ and exponents $p_\gamma$ 
is given in Table~1. For example, the $A_1A_1$ and $A_1A_2$ 
amplitudes read
\bea
&& S_{11}(\theta)=t_{2/3}(\theta)t_{2/5}(\theta)t_{1/15}(\theta)
\label{S11}\\
&& S_{12}(\theta)=t_{4/5}(\theta)t_{3/5}(\theta)t_{7/15}(\theta)
t_{4/15}(\theta)\,\,.
\label{S12}
\eea
Equation (\ref{triangle}) shows that the particles $A_1$, $A_2$, $A_3$ ($A_1$, 
$A_2$, $A_3$, $A_4$) appear as bound states in the $A_1A_1$ ($A_1A_2$) 
scattering channel.

\begin{table}
\begin{center}
\begin{tabular}{|c|l|}\hline
$a$ \,\, $b$ &
$S_{ab}$ \\ \hline 
1 \,\, 1 &
$ \st{\bf 1}{(20)} \, \st{\bf 2}{(12)} \, \st{\bf 3}{(2)} $\\ 
1 \,\, 2 &
$ \st{\bf 1}{(24)} \, \st{\bf 2}{(18)} \, \st{\bf 3}{(14)} \, \st{\bf 4}{(8)}
$\\ 
1 \,\, 3 &
$ \st{\bf 1}{(29)} \, \st{\bf 2}{(21)} \, \st{\bf 4}{(13)} \,
\st{\bf 5}{(3)} \, (11)^2 $ \\ 
1 \,\, 4 &
$ \st{\bf 2}{(25)} \, \st{\bf 3}{(21)} \, \st{\bf 4}{(17)} \,
\st{\bf 5}{(11)} \, \st{\bf 6}{(7)} \, (15) $ \\ 
1 \,\, 5 &
$ \st{\bf 3}{(28)} \, \st{\bf 4}{(22)} \, \st{\bf 6}{(14)} \,
\st{\bf 7}{(4)} \, (10)^2 \, (12)^2 $ \\ 
1 \,\, 6 &
$ \st{\bf 4}{(25)} \, \st{\bf 5}{(19)} \, \st{\bf 7}{(9)} \,
(7)^2 \, (13)^2 \, (15) $ \\ 
1 \,\, 7 &
$ \st{\bf 5}{(27)} \, \st{\bf 6}{(23)} \, \st{\bf 8}{(5)} \,
(9)^2 \, (11)^2\, (13)^2 \, (15) $ \\ 
1 \,\, 8 &
$ \st{\bf 7}{(26)} \, \st{\bf 8}{(16)^3} \, (6)^2 \,
(8)^2 \, (10)^2 \, (12)^2 $ \\ 
2 \,\, 2 &
$ \st{\bf 1}{(24)} \, \st{\bf 2}{(20)} \, \st{\bf 4}{(14)} \,
\st{\bf 5}{(8)} \,\st{\bf 6}{(2)} \, (12)^2 $ \\ 
2 \,\, 3 &
$ \st{\bf 1}{(25)} \, \st{\bf 3}{(19)} \, \st{\bf 6}{(9)} \, (7)^2 \,
(13)^2 \, (15) $ \\ 
2 \,\, 4 &
$ \st{\bf 1}{(27)} \,\st{\bf 2}{(23)} \, \st{\bf 7}{(5)} \,
(9)^2 \, (11)^2 \, (13)^2\,(15)$ \\ 
2 \,\, 5 &
$ \st{\bf 2}{(26)} \,\st{\bf 6}{(16)^3} \, (6)^2 (8)^2 (10)^2 (12)^2 $
\\ 
2 \,\, 6 &
$ \st{\bf 2}{(29)} \, \st{\bf 3}{(25)} \, \st{\bf 5}{(19)^3} \,
\st{\bf 7}{(13)^3} \, \st{\bf 8}{(3)} \, (7)^2 (9)^2 (15) $ \\ 
2 \,\, 7 &
$ \st{\bf 4}{(27)} \, \st{\bf 6}{(21)^3} \, \st{\bf 7}{(17)^3} \,
\st{\bf 8}{(11)^3} \, (5)^2 (7)^2 (15)^2 $ \\ 
2 \,\, 8 &
$ \st{\bf 6}{(28)} \, \st{\bf 7}{(22)^3} \, (4)^2 (6)^2 (10)^4 (12)^4 (16)^4 $
\\ 
3 \,\, 3 &
$ \st{\bf 2}{(22)} \, \st{\bf 3}{(20)^3} \, \st{\bf 5}{(14)} \,
\st{\bf 6}{(12)^3} \, \st{\bf 7}{(4)} \, (2)^2 $ \\ 
3 \,\, 4 &
$ \st{\bf 1}{(26)} \, \st{\bf 5}{(16)^3} \, (6)^2 (8)^2 (10)^2 (12)^2 $
\\ 
3 \,\, 5 &
$ \st{\bf 1}{(29)} \, \st{\bf 3}{(23)} \, \st{\bf 4}{(21)^3} \,
\st{\bf 7}{(13)^3} \, \st{\bf 8}{(5)} \, (3)^2 (11)^4 (15) $ \\ 
3 \,\, 6 &
$ \st{\bf 2}{(26)} \, \st{\bf 3}{(24)^3} \, \st{\bf 6}{(18)^3} \,
\st{\bf 8}{(8)^3} \, (10)^2 (16)^4 $ \\ 
3 \,\, 7 &
$ \st{\bf 3}{(28)} \, \st{\bf 5}{(22)^3} \, (4)^2 (6)^2 (10)^4 (12)^4 (16)^4
$ \\ 
3 \,\, 8 &
$ \st{\bf 5}{(27)} \, \st{\bf 6}{(25)^3} \, \st{\bf 8}{(17)^5} \,
(7)^4 (9)^4 (11)^2 (15)^3 $ \\ 
4 \,\, 4 &
$ \st{\bf 1}{(26)} \, \st{\bf 4}{(20)^3} \, \st{\bf 6}{(16)^3} \,
\st{\bf 7}{(12)^3} \, \st{\bf 8}{(2)} \, (6)^2 (8)^2 $ \\ 
4 \,\, 5 &
$ \st{\bf 1}{(27)} \, \st{\bf 3}{(23)^3} \, \st{\bf 5}{(19)^3} \,
\st{\bf 8}{(9)^3} \, (5)^2 (13)^4 (15)^2 $ \\ 
4 \,\, 6 &
$ \st{\bf 1}{(28)} \, \st{\bf 4}{(22)^3} (4)^2 (6)^2 (10)^4
(12)^4 (16)^4 $ \\ 
4 \,\, 7 &
$ \st{\bf 2}{(28)} \, \st{\bf 4}{(24)^3} \, \st{\bf 7}{(18)^5} \,
\st{\bf 8}{(14)^5} \, (4)^2 (8)^4 (10)^4 $ \\ 
4 \,\, 8 &
$ \st{\bf 4}{(29)} \, \st{\bf 5}{(25)^3} \, \st{\bf 7}{(21)^5} \,
(3)^2 (7)^4 (11)^6 (13)^6 (15)^3 $ \\ 
\hline
\end{tabular}
\end{center}
\end{table}

\newpage
\begin{table}
\begin{center}
\begin{tabular}{|c|l|}\hline
5 \,\, 5 &
$ \st{\bf 4}{(22)^3} \, \st{\bf 5}{(20)^5} \, \st{\bf 8}{(12)^5} \,
(2)^2 (4)^2 (6)^2 (16)^4 $ \\ 
5 \,\, 6 &
$ \st{\bf 1}{(27)} \, \st{\bf 2}{(25)^3} \, \st{\bf 7}{(17)^5} \,
(7)^4 (9)^4 (11)^4 (15)^3 $ \\ 
5 \,\, 7 &
$ \st{\bf 1}{(29)} \, \st{\bf 3}{(25)^3} \, \st{\bf 6}{(21)^5} \,
(3)^2 (7)^4 (11)^6 (13)^6 (15)^3 $ \\ 
5 \,\, 8 &
$ \st{\bf 3}{(28)} \, \st{\bf 4}{(26)^3} \, \st{\bf 5}{(24)^5} \,
\st{\bf 8}{(18)^7} \, (8)^6 (10)^6 (16)^8 $ \\ 
6 \,\, 6 &
$ \st{\bf 3}{(24)^3} \, \st{\bf 6}{(20)^5} \, \st{\bf 8}{(14)^5} \,
(2)^2 (4)^2 (8)^4 (12)^6 $ \\ 
6 \,\, 7 &
$ \st{\bf 1}{(28)} \, \st{\bf 2}{(26)^3} \, \st{\bf 5}{(22)^5} \,
\st{\bf 8}{(16)^7} \, (6)^4 (10)^6 (12)^6 $ \\ 
6 \,\, 8 &
$ \st{\bf 2}{(29)} \, \st{\bf 3}{(27)^3} \, \st{\bf 6}{(23)^5} \,
\st{\bf 7}{(21)^7} \, (5)^4 (11)^8 (13)^8 (15)^4 $ \\ 
7 \,\, 7 &
$ \st{\bf 2}{(26)^3} \, \st{\bf 4}{(24)^5} \, \st{\bf 7}{(20)^7} \,
(2)^2 (8)^6 (12)^8 (16)^8 $ \\ 
7 \,\, 8 &
$ \st{\bf 1}{(29)} \, \st{\bf 2}{(27)^3} \, \st{\bf 4}{(25)^5} \,
\st{\bf 6}{(23)^7} \, \st{\bf 8}{(19)^9} \, (9)^8 (13)^{10} (15)^ 5 $
\\ 
8 \,\, 8 &
$ \st{\bf 1}{(28)^3} \, \st{\bf 3}{(26)^5} \, \st{\bf 5}{(24)^7} \,
\st{\bf 7}{(22)^9} \, \st{\bf 8}{(20)^{11}} \, (12)^{12} (16)^{12} $ \\
\hline
\end{tabular}
\caption{$S$-matrix of the scaling Ising model in a 
magnetic field at $T=T_c$. A factor $\left(t_{\gamma/30}(\theta)
\right)^{p_\gamma}$ in the amplitude $S_{ab}(\theta)$ corresponds to each term 
$(\gamma)^{p_\gamma}$ ($p_\gamma=1$ is 
omitted; the building blocks $t_\alpha(\theta)$ are defined in (\ref{talpha})).
The superscript {\bf c} above $(\gamma)$ indicates that the pole at $\theta=
i\pi\gamma/30$ in the amplitude $S_{ab}(\theta)$ corresponds to the 
particle $A_c$ appearing as bound state in the $A_aA_b$ scattering channel.}
\end{center}
\end{table}

Most of the amplitudes in Table~1 contain higher order poles that we did not 
discuss in the previous section. Within the framework of the analytic 
$S$-matrix, each singularity is expected to have a physical interpretation.
The higher order poles of the scattering amplitudes were recognised in 
\cite{CT} (see also \cite{BCDS,ChM}) as the singularities that in two 
dimensions are associated to processes in which more than one particle 
propagates in the intermediate state\footnote{In four dimensions these 
processes give rise to anomalous thresholds rather than poles \cite{ELOP}.}.
Figures~7 and 8 show the processes of this kind leading to second and third 
order poles at $\theta=i\varphi$,
\EQ
\varphi=u_{ad}^c+u_{db}^e-\pi\,,
\label{phi}
\EN
in the amplitude $S_{ab}(\theta)$. The angle $\eta$ in Figure~7a is
\EQ
\eta=\pi-u^a_{cd}-u^b_{de}\in[0,\pi)\,,
\EN
and $i\eta$ is the rapidity difference between the intermediate propagating 
particles $A_c$ and $A_e$; the diagram of Figure~7b corresponds to 
the limiting case $\eta=0$. One has
\EQ
S_{ab}(\theta\simeq i\varphi) \,\simeq\,
\frac{(\Gamma_{cd}^a \Gamma_{de}^b)^2S_{ce}(i\eta)}{(\theta-i\varphi)^2}
\lab{res2a}
\EN
\EQ
S_{ab}(\theta\simeq i\varphi) \,\simeq\,
\frac{\Gamma_{cd}^a \Gamma_{de}^b\Gamma_{cf}^b\Gamma_{fe}^a}
{(\theta-i\varphi)^2}
\lab{res2b}
\EN
for the diagrams of Figures~7a and 7b, respectively, and 
\EQ
S_{ab}(\theta\simeq i\varphi)\simeq i \,
\frac{(\Gamma_{cd}^a \Gamma_{de}^b \Gamma_{ec}^f)^2}{(\theta-i\varphi)^3}
\lab{res3}
\EN
for the direct channel third order poles of Figure~8. More generally, a pole of
order $P-2L$ corresponds to a diagram with $P$ internal lines and $L$ loops 
\cite{BCDS}.

\begin{figure}
\centerline{
\includegraphics[width=9cm]{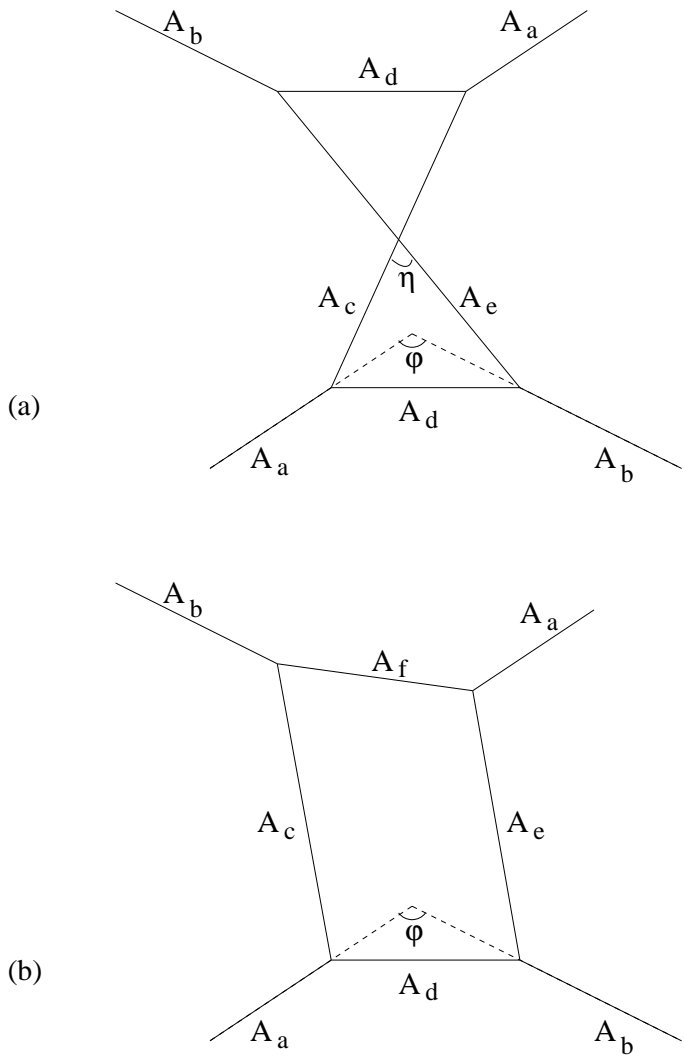}}
\caption{Scattering patterns associated to the second order poles of the 
$S$-matrix.}
\end{figure}

\begin{figure}
\centerline{
\includegraphics[width=9cm]{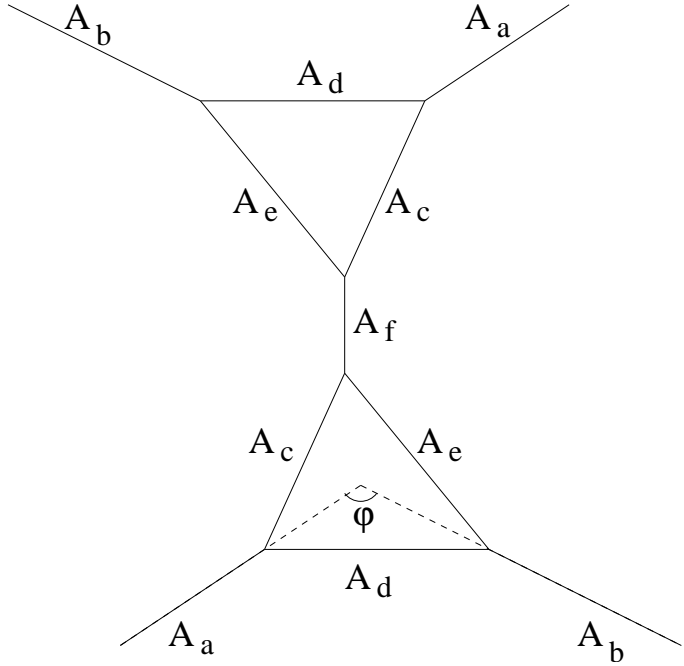}}
\caption{Third order pole of the scattering amplitudes.}
\end{figure}

We said that the $S$-matrix of Table~1 is the minimal solution to the 
constraints discussed so far. Although it is extremely natural to expect that 
this minimal $S$-matrix is the one describing the magnetic Ising
model at $T=T_c$, the conjecture needs to be checked. Al.~Zamolodchikov 
showed that the conformal limit of an integrable field theory can be identified
using the $S$-matrix as the only imput. The method, known as thermodynamic
Bethe ansatz, allows the computation of the central 
charge of the conformal limit through the study of the thermodynamics on a 
cylindrical geometry \cite{TBA}. When applied to the $S$-matrix of Table~1 it
yields the expected result $C=1/2$ \cite{KM}. 

A different way of confirming that the $S$-matrix of Table~1 does correspond
to the Ising model is that of showing that it describes an integrable quantum
field theory containing two relevant scalar operators besides the 
identity\footnote{This requirement uniquely identifies the Ising model among
the theories satisfying reflection positivity.}. Addressing this problem
means taking Zamolodchikov's $S$-matrix as the imput of the form factor 
equations (\ref{ff0}--\ref{ff4}) and looking for scalar solutions which 
behave asymptotically as (\ref{asymptotics}) with $Y_\Phi<1$ \cite{immf,DS} . 
The non-degenerate spectrum and the ``$\varphi^3$-property'' $\Gamma_{11}^1\neq
0$ show that the theory possesses no internal symmetries, so that all the form
factors of scalar primaries other than the identity are expected to be 
non-vanishing. For operators local with respect to the particles ($l_{\Phi,
\phi_a}=1$), the general solution of equations (\ref{ff1}) and (\ref{ff3}) 
with the required pole structure can be written in the form 
\EQ
F^\Phi_{a_1\ldots a_n}(\theta_1,\ldots,\theta_n)=
Q^\Phi_{a_1\ldots a_n}(\theta_1,\ldots,\theta_n)\prod_{i<j}\frac{
F^{min}_{a_ia_j}(\theta_i-\theta_j)}{\left(\cosh\left(\frac{\theta_i-
\theta_j}{2}\right)\right)^{\delta_{a_ia_j}}D_{a_ia_j}(\theta_i-\theta_j)}\,\,.
\label{fn}
\EN
Here $F^{min}_{ab}(\theta)$ is a solution of the equations 
\EQ
F(\theta)=S_{ab}(\theta)F(-\theta)
\EN
\EQ
F(\theta+2i\pi)=F(-\theta)
\EN
free of zeroes and poles for Im$\theta\in(0,2\pi)$. In the denominator, the 
factors $\cosh\left(\frac{\theta_i-\theta_j}{2}\right)$ introduce the 
annihilation poles, 
while $D_{ab}(\theta)$ takes care of the dynamical poles in the $A_aA_b$ 
scattering channel through factors of the type $(\cosh\theta-\cos u_{ab}^c)$ 
(see later for the precise form). Finally, the $Q^\Phi_{a_1\ldots a_n}$ are
entire functions of the rapidities, invariant under exchanges 
$\theta_i\leftrightarrow\theta_j$ and (up to a sign) $2\pi i$-periodic in all 
$\theta_j$'s. They are subject to (\ref{ff0}) and to the residue equations
(\ref{ff2}) and 
(\ref{ff4}) which relate functions with different $n$. These equations,
however, hold for any operator with a given spin, so that further constraints 
are needed to select specific solutions. We will see in a moment the role 
played in this respect by the asymptotic bound (\ref{bound}).

The form factors are determined starting with the first non-trivial case
(the two-particle one) and then using the residue equations to fix the 
matrix elements with an higher number of particles (form factor bootstrap). 
The specialisation of (\ref{fn}) to $n=2$ and scalar operators reads
\EQ
F^\Phi_{ab}(\theta)=\frac{Q_{ab}^\Phi(\theta)}{D_{ab}(\theta)}\,
F^{min}_{ab}(\theta)\,,
\label{f2}
\EN
where we made the identifications $F^\Phi_{ab}(\theta_1,\theta_2)\equiv
F^\Phi_{ab}(\theta_1-\theta_2)$, $Q^\Phi_{ab}(\theta_1,\theta_2)\equiv
Q^\Phi_{ab}(\theta_1-\theta_2)$, and took into account the vanishing residue
on the annihilation pole in the two-particle case when $l_{\Phi,\phi_a}=1$.

The functions $F^{min}_{ab}(\theta)$ with the properties specified above can
be written as
\EQ
F^{min}_{ab}(\theta)=\left(-i\sinh\frac{\theta}{2}\right)^{\delta_{ab}}
\prod_{\gamma\in{\cal G}_{ab}}\left(T_{\gamma/30}(\theta)
\right)^{p_\gamma}\,\,,
\lab{fmin}
\EN
where
\EQ
T_{\alpha}(\theta)=\exp\left\{2\int_0^\infty\frac{dt}{t}\frac{\cosh\left(
\alpha - \frac{1}{2}\right)t}{\cosh\frac{t}{2}\sinh
t}\sin^2\frac{(i\pi-\theta)t}{2\pi}\right\}
\lab{Talpha}
\EN
solves the equations
\EQ
T_\alpha(\theta)=-t_\alpha(\theta)T_\alpha(-\theta)
\EN
\EQ
T_\alpha(\theta+2i\pi)=T_\alpha(-\theta)\,;
\EN
the property
\EQ
S_{ab}(0)=(-1)^{\delta_{ab}}
\EN
of the $S$-matrix originates the factor in front of the product in 
(\ref{fmin}).

As we said, the denominator of (\ref{f2}) takes care of the dynamical poles.
While equation (\ref{ff2}) says how to deal with the simple poles of the 
$S$-matrix, the case of higher order poles needs to be discussed  
\cite{immf}. Due to crossing symmetry, poles appear in pairs in the scattering
amplitudes. For poles of even order both residues are positive and there is
no way to distinguish between a direct and a crossed channel. Then a diagram of
the type shown in Figure~7 must exist for each double pole at 
$\theta=i\varphi$ in the amplitude $S_{ab}(\theta)$. In the vicinity of such a 
pole, the form factor behaves as\footnote{The pole in the form factor is simple
because the diagram of Figure~9 has a single triangular loop.} (see 
Figure~9)
\EQ
F_{ab}^{\Phi}(\theta\simeq i\varphi) \simeq
i \,\frac{\Gamma_{cd}^a \Gamma_{de}^b S_{ce}(i\eta) F_{ce}^{\Phi}(-i\eta)}
{\theta-i\varphi}=i \,\frac{\Gamma_{cd}^a \Gamma_{de}^b F_{ce}^{\Phi}(i\eta)}
{\theta-i\varphi}
\lab{ffres2}
\EN
(this result also holds for $\eta=0$).

A third order pole in the scattering amplitude can be seen as originating from 
a double pole when $\eta=u^f_{ce}$. Then the direct channel singularity at 
$\theta=i\varphi$ in the form factor is obtained using (\ref{ff2}) into 
(\ref{ffres2}) and reads (Figure~10) 
\EQ
F_{ab}^{\Phi}(\theta\simeq i\varphi)\simeq -\frac{\Gamma_{cd}^a \Gamma_{de}^b
\Gamma_{ec}^f}{(\theta-i\varphi)^2}F_f^{\Phi}\,,
\lab{ffres3}
\EN
while the crossed channel pole at $\theta=i(\pi-\varphi)$ remains simple.

\begin{figure}
\centerline{
\includegraphics[width=12.5cm]{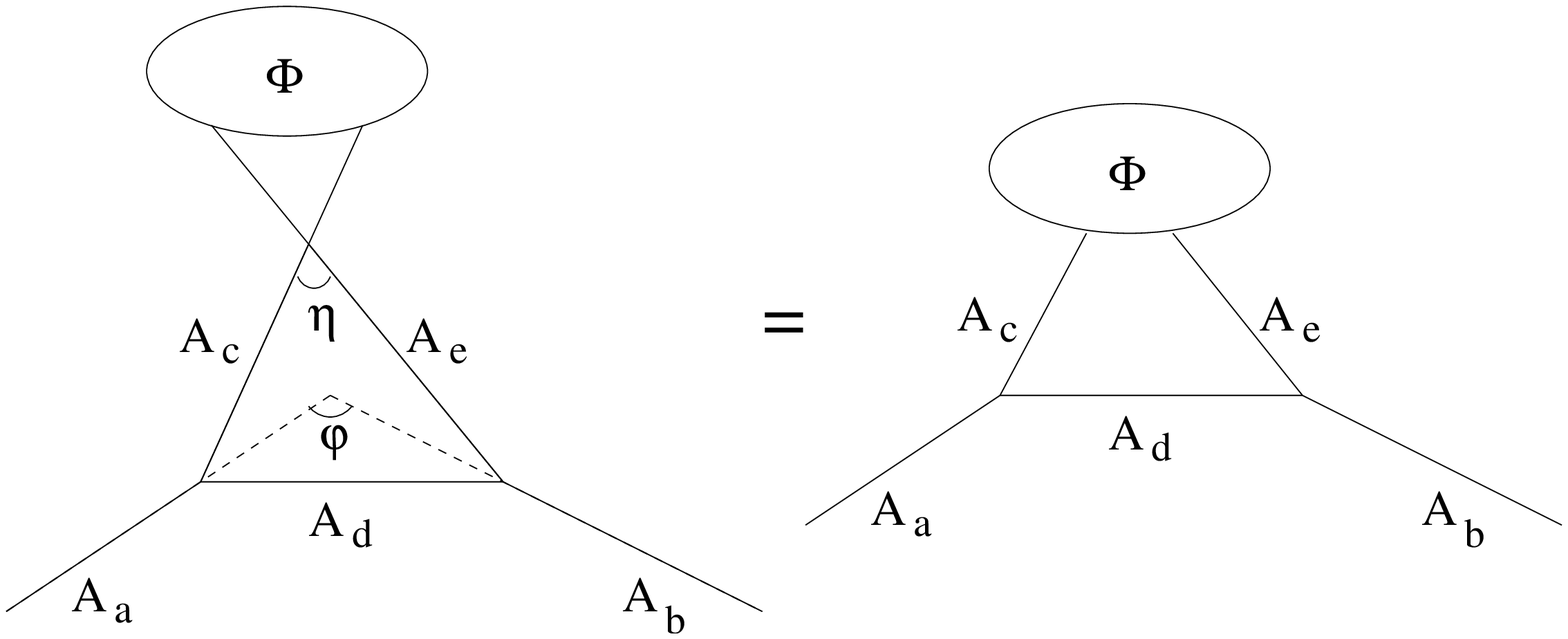}}
\caption{Form factor singularity associated to a double pole of the 
$S$-matrix.}
\vspace{1cm}
\centerline{
\includegraphics[width=6.5cm]{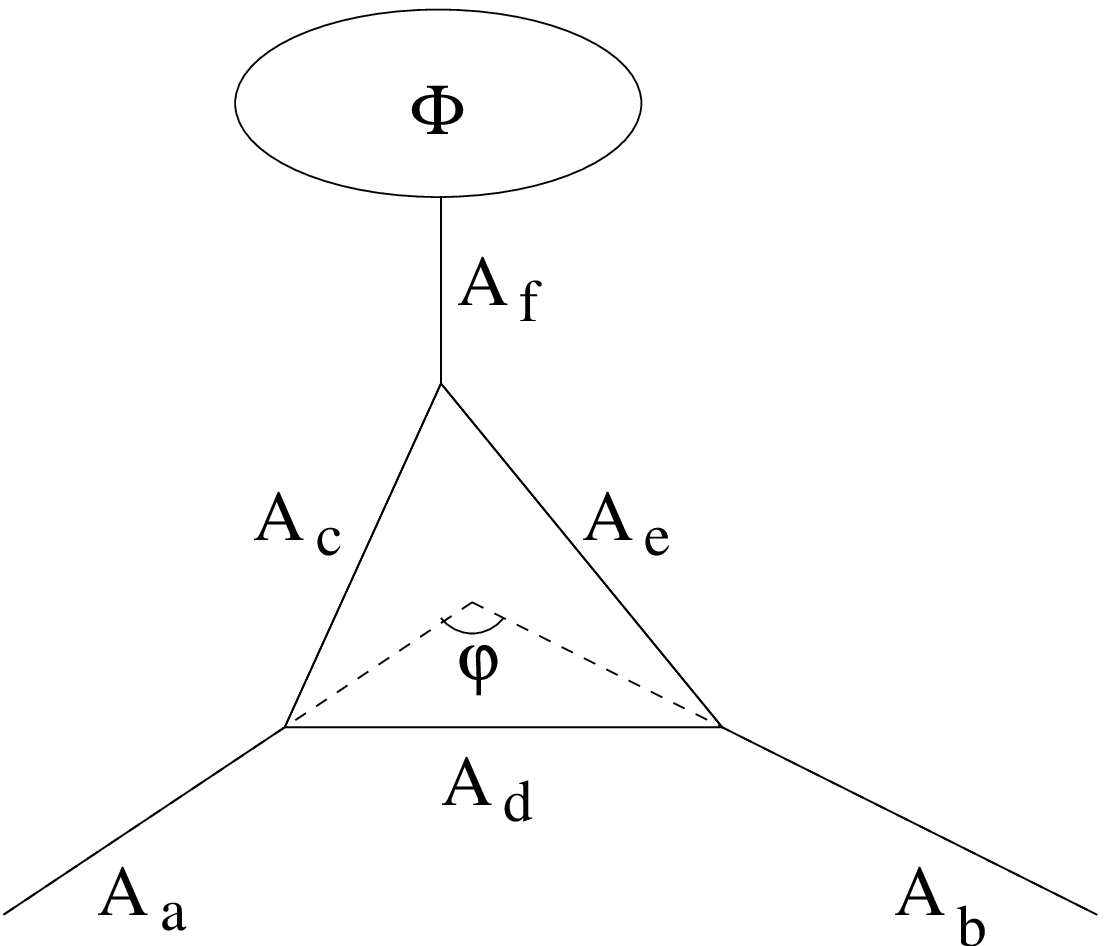}}
\caption{Form factor singularity associated to a direct channel triple pole 
of the $S$-matrix.}
\end{figure}

The analysis of poles of higher order can be done along similar lines and 
leads to the expression
\EQ
D_{ab}(\theta)=\prod_{\gamma\in{\cal G}_{ab}}\left({\cal P}_{\gamma/30}(\theta)
\right)^{i_\gamma}\left({\cal P}_{1-\gamma/30}(\theta)\right)^{j_\gamma}\,,
\lab{dab}
\EN
where
\EQ
{\cal P}_{\alpha}(\theta)=\frac{\cos\pi\alpha-\cosh\theta}
{2\cos^2\frac{\pi\alpha}{2}}
\lab{polo}
\EN
and
\EQ
\begin{array}{lll}
i_{\gamma} = n+1\,,\hspace{.5cm}& j_{\gamma} = n \,, &
\hspace{.5cm}{\rm if}\hspace{.3cm} p_\gamma=2n+1\\
i_{\gamma} = n \,,\hspace{.5cm}& j_{\gamma} = n \,, &
\hspace{.5cm}{\rm if}\hspace{.3cm} p_\gamma=2n\,\,.
\end{array}
\EN

We are now in the position of dealing with the functions $Q^\Phi_{ab}(\theta)$,
the only piece of (\ref{f2}) which carries information about the operators.
As functions of the rapidity difference, they must be even, $2\pi i$-periodic,
free of poles and exponentially bounded. Hence we can write them as
\EQ
Q_{ab}^\Phi(\theta)=\sum_{k=0}^{N_{ab}^\Phi} c^{(k)}_{ab,\Phi}\,
\cosh^k\theta\,\,.
\label{qab}
\EN
The condition
\EQ
\left[F^\Phi_{ab}(\theta)\right]^*=F^\Phi_{ab}(-\theta)
\EN
follows from (\ref{ff1}) and $S_{ab}^*(\theta)=S_{ab}(-\theta)$, and ensures 
that the coefficients $c^{(k)}_{ab,\Phi}$ are real. These coefficients are the 
only unknowns we are left with.

Let us look for solutions corresponding to {\em relevant} operators. This means
that the asymptotic behaviour (\ref{asymptotics}) has to hold with $Y_\Phi<1$.
Since
\EQ
T_{\alpha}(\theta)\sim\exp(|\theta|/2)\,,\hspace{1cm}|\theta|\goto\infty\,,
\EN
it is straightforward to check on (\ref{f2}) that, in particular,
\EQ
N_{11}^{\Phi}\leq 1\,\,.
\EN
Hence, the initial condition of the form factor bootstrap for relevant scalar
operators allows for {\em two} free parameters, i.e. the coefficients
$c_{11,\Phi}^{(0)}$ and $c_{11,\Phi}^{(1)}$. It can be checked that the number
of free parameters does not increase when implementing the bootstrap. For 
example, since 
\EQ
N_{12}^{\Phi}\leq 2\,,
\EN
considering $F^\Phi_{12}(\theta)$ brings in three more coefficients. On the 
other hand, the amplitudes $S_{11}(\theta)$ and $S_{12}(\theta)$ 
possess three common bound states, what yields the three equations
\EQ
\frac{1}{\Gamma_{11}^c}\mbox{Res}_{\theta=iu_{11}^c}F_{11}^{\Phi}(\theta)=
\frac{1}{\Gamma_{12}^c}\mbox{Res}_{\theta=iu_{12}^c}F_{12}^{\Phi}(\theta)\,,
\hspace{1cm}c=1,2,3
\EN
which determine the three $c^{(k)}_{12,\Phi}$ in terms of the two 
$c^{(k)}_{11,\Phi}$.
Going on and using also the conditions on higher order poles, a number of 
residue equations larger than the number of new coefficients is available in 
many cases. It turns out, however, that the extra constraints are automatically
fulfilled so that the two initial parameters propagate untouched through the 
bootstrap procedure \cite{immf,DS}.  

Since all the form factor equations used so far are linear in the operator
$\Phi$, the interpretation of this result is simple: the Zamolodchikov's 
$S$-matrix describes an integrable field theory with two {\em scaling} relevant
scalar operators $\Phi_1$ and $\Phi_2$ (plus the identity) and, up to an
additive constant, the operators $\Phi$ selected up to now can be written as
\EQ
\Phi=\alpha\,\Phi_1+\beta\,\Phi_2
\label{linear}
\EN
with $\alpha$ and $\beta$ real parameters. We already said that this condition
uniquely identifies the Ising field theory (\ref{A}) in which the two scaling 
operators are $\sigma$ and $\varepsilon$; the known results about the 
integrable directions imply $\tau=0$, $h\neq 0$. Hence, the analysis of the 
operator content confirms that the Zamolodchikov's $S$-matrix corresponds to 
the integrable direction of the magnetic Ising model.

Obviously, the next issue is that of identifying the solutions 
corresponding to the scaling operators $\sigma$ and $\varepsilon$. Having 
already exploited all the constraints on form factors discussed so far, it is 
clear that this task requires some new physical input. Since the scaling 
operators are characterised by their behaviour close to criticality, it is 
natural to look once again at the asymptotic properties of form factors. 
More precisely, we consider the limit \cite{DSC}
\EQ
\lim_{\alpha\goto+\infty}F^{\hat{\Phi}_k}_{a_1\ldots a_rb_1\ldots b_{l}}
(\theta_1+\frac\alpha2,\ldots,\theta_r+\frac\alpha2,\theta_1'-\frac\alpha2,
\ldots,\theta_l'-\frac\alpha2)\,,
\label{massless}
\EN
where
\EQ
\hat{\Phi}_k=\frac{\Phi_k}{\langle\Phi_k\rangle}\,,\hspace{1cm}k=1,2
\label{phihat}
\EN
are the scaling operators entering (\ref{linear}) rescaled by their vacuum 
expectation value. It follows from (\ref{onshell}) that shifting a rapidity 
$\theta$ by $\pm\alpha/2$ and rescaling the mass as $m_a=M_ae^{-\alpha/2}$
produces, when $\alpha\goto+\infty$, a massless particle with energy
$p^0=\pm p^1=\frac{M_a}{2}e^{\pm\theta}$ (right- or left-mover, depending on 
the sign of momentum). Hence, (\ref{massless}) corresponds to the massless
limit towards the conformal point in which the first $r$ particles become
right-movers and the remaining $l$ particles left-movers. The property
\EQ
\lim_{\alpha\goto+\infty}S_{ab}(\theta+\alpha)=1
\EN
of the scattering amplitudes of Table~1 shows that a right-mover and a 
left-mover do not interact in such a conformal limit. 
We conclude that
the limit (\ref{massless}) produces a factorisation into two massless form 
factors, one with $r$ right-movers and one with $l$ left-movers. On the other
hand, a massless form factor with all right (left) movers is obtained from
(\ref{massless}) with $l=0$ ($r=0$); in such a case all rapidities are shifted 
by the same amount, and (\ref{ff0}) shows that this massless form factor 
actually coincides with the massive one. Then we conclude that the 
asymptotic factorisation property
\EQ
\lim_{\alpha\goto+\infty}F^{\hat{\Phi}_k}_{a_1\ldots a_rb_1\ldots b_{l}}
(\theta_1+\alpha,\ldots,\theta_r+\alpha,\theta_1',\ldots,\theta_l')=
F^{\hat{\Phi}_k}_{a_1\ldots a_{r}}(\theta_1,\ldots,\theta_r)
F^{\hat{\Phi}_k}_{b_{1}\ldots b_{l}}(\theta_1',\ldots,\theta_l')
\label{cluster}
\EN
holds in the original massive theory. Notice that the particular cases $r=0$ 
and/or $l=0$ require the normalisation (\ref{phihat}). The factorisation
argument leading to (\ref{cluster}) applies to scaling operators and not to 
linear combinations like (\ref{linear}) mixing operators with different scaling
dimensions: in the latter case one of the coefficient $\alpha$, $\beta$ is 
dimensionful and vanishes in the massless limit.

We are now in the position of identifying the initial conditions of the form
factor bootstrap corresponding to the scaling operators $\Phi_1$ and $\Phi_2$.
This amounts to finding the values of the ratio
\EQ
z_{\Phi}=\frac{c_{11,\Phi}^{(0)}}{c_{11,\Phi}^{(1)}}
\label{z}
\EN
for which the factorisation conditions (\ref{cluster}) are fulfilled (this 
quantity is universal because does not depend on the normalisation of the 
operator). The equation
\EQ
\frac{1}{F_1^{\Phi_k}}\lim_{\theta\rightarrow\infty}
F_{11}^{\Phi_k}(\theta)=\frac{1}{F_2^{\Phi_k}}\lim_{\theta\rightarrow\infty}
F_{12}^{\Phi_k}(\theta)
\EN
is a consequence of (\ref{cluster}) and gives a quadratic equation for 
$z_\Phi$ whose solutions are \cite{DS}
\bea
&& z_{\Phi_1}=4.869840..\nonumber\\
&& z_{\Phi_2}=1.255585..\,\,.\nonumber
\eea

Being the operator which perturbs the conformal point, $\sigma$ is proportional
to the trace of the energy-momentum tensor. It is not difficult to see that
the conservation of the latter induces the factorisation of a kinematical
term in the form factors of $\sigma$. In particular, the two-particle form 
factors $F^\sigma_{ab}(\theta)$ contain the factor \cite{immf}
\EQ
\left(\cosh\theta+\frac{m_a^2+m_b^2}{2m_am_b}\right)^{1-\delta_{ab}}\,\,.
\EN
It can be checked that the form factors originating from the initial condition
$z_{\Phi_1}$ have this property. Then we conclude
\bea
&& \Phi_1=\sigma\nonumber\\
&& \Phi_2=\varepsilon\nonumber\,\,.
\eea

\begin{table}
\begin{center}
\begin{tabular}{|c|r|r|}\hline
   & $\hat{\sigma}$\hspace{1cm} & $\hat{\var}$\hspace{1cm} \\ \hline\hline
$ c_{11}^1 $ & $ -2.093102832 $ & $ -70.00917205 $ \\       
$ c_{11}^0 $ & $ -10.19307727 $ & $ -87.90247670 $ \\ \hline
$ c_{12}^2 $ & $ -7.979022182 $ & $ -466.3008246 $ \\
$ c_{12}^1 $ & $ -71.79206351 $ & $ -1307.331521 $ \\
$ c_{12}^0 $ & $ -70.29218939 $ & $ -853.2803886 $ \\ \hline
$ c_{13}^3 $ & $ -582.2557366 $ & $ -43021.45153 $ \\
$ c_{13}^2 $ & $ -6944.416956 $ & $ -182413.2733 $ \\
$ c_{13}^1 $ & $ -13406.48877 $ & $ -241929.7678 $ \\
$ c_{13}^0 $ & $ -7049.622303 $ & $ -102574.1349 $ \\ \hline
$ c_{22}^3 $ & $ -21.48559881 $ & $ -2193.896354 $ \\
$ c_{22}^2 $ & $ -333.8125724 $ & $ -10870.05277 $ \\
$ c_{22}^1 $ & $ -791.3745549 $ & $ -16161.44508 $ \\
$ c_{22}^0 $ & $ -500.2535896 $ & $ -7510.235388 $ \\ \hline
$ c_{14}^3 $ & $ 22.57778351 $ & $ 2074.636471 $ \\
$ c_{14}^2 $ & $ 318.7122159 $ & $ 9881.413381 $ \\
$ c_{14}^1 $ & $ 672.2210098 $ & $ 14357.04570 $ \\
$ c_{14}^0 $ & $ 377.4586311 $ & $ 6568.762583 $ \\ \hline
$ c_{15}^4 $ & $ -260.7643072 $ & $ -30333.56619 $ \\
$ c_{15}^3 $ & $ -4719.877128 $ & $ -198757.2340 $ \\
$ c_{15}^2 $ & $ -15172.07643 $ & $ -447504.5720 $ \\
$ c_{15}^1 $ & $ -17428.22924 $ & $ -422808.9295 $ \\
$ c_{15}^0 $ & $ -6716.787925 $ & $ -143743.2050 $ \\ \hline
$ c_{23}^4 $ & $ -92.73452350 $ & $ -11971.94909 $ \\
$ c_{23}^3 $ & $ -1846.579035 $ & $ -81253.72269 $ \\
$ c_{23}^2 $ & $ -6618.297073 $ & $ -186593.8661 $ \\
$ c_{23}^1 $ & $ -8436.850082 $ & $ -178494.3378 $ \\
$ c_{23}^0 $ & $ -3579.556465 $ & $ -61194.62416 $ \\ \hline
$ c_{33}^5 $ & $ -1197.056497 $ & $ -195385.7662 $ \\
$ c_{33}^4 $ & $ -30166.99117 $ & $ -1743171.802 $ \\
$ c_{33}^3 $ & $ -150512.4122 $ & $ -5603957.324 $ \\
$ c_{33}^2 $ & $ -301093.9432 $ & $ -8422606.859 $ \\
$ c_{33}^1 $ & $ -267341.1276 $ & $ -6035102.896 $ \\
$ c_{33}^0 $ & $ -87821.70785 $ & $ -1668721.004 $ \\ \hline
\end{tabular}
\end{center}
\end{table}

\begin{table}
\begin{center}
\begin{tabular}{|c|r|r|}\hline
$ c_{25}^6 $ & $ 1425.995027 $ & $ 289831.4882 $ \\
$ c_{25}^5 $ & $ 44219.03877 $ & $ 3275586.983 $ \\
$ c_{25}^4 $ & $ 286184.1535 $ & $ 13872077.63 $ \\
$ c_{25}^3 $ & $ 788413.2178 $ & $ 29236961.96 $ \\
$ c_{25}^2 $ & $ 1078996.488 $ & $ 32979257.31 $ \\
$ c_{25}^1 $ & $ 725356.4417 $ & $ 19100224.04 $ \\
$ c_{25}^0 $ & $ 191383.5734 $ & $ 4471623.121 $ \\ \hline
$ c_{17}^5 $ & $ 190.8548023 $ & $ 30394.23374 $ \\
$ c_{17}^4 $ & $ 4633.706068 $ & $ 274294.8033 $ \\
$ c_{17}^3 $ & $ 21406.72691 $ & $ 897781.3229 $ \\
$ c_{17}^2 $ & $ 39514.82959 $ & $ 1375919.456 $ \\
$ c_{17}^1 $ & $ 32456.91939 $ & $ 1004969.466 $ \\
$ c_{17}^0 $ & $ 9906.265607 $ & $ 282938.1974 $ \\ \hline
$ c_{44}^7 $ & $ -7249.785565 $ & $ -1830120.693 $ \\
$ c_{44}^6 $ & $ -276406.7236 $ & $ -25699492.93 $ \\
$ c_{44}^5 $ & $ -2299573.212 $ & $ -138411873.8 $ \\
$ c_{44}^4 $ & $ -849276.3526 $ & $ -384776478.8 $ \\
$ c_{44}^3 $ & $ -16615618.39 $ & $ -608371427.1 $ \\
$ c_{44}^2 $ & $ -17950817.11 $ & $ -553818699.0 $ \\
$ c_{44}^1 $ & $ -10139089.36 $ & $ -270964337.7 $ \\
$ c_{44}^0 $ & $ -2341590.241 $ & $ -55283137.91 $ \\ \hline
\end{tabular}
\caption{Coefficients of some of the polynomials (\ref{qab}) for the
primary operators $\hat{\sigma}$ and $\hat{\varepsilon}$ (parenthesis on the 
upper index of the coefficients are omitted; $\hat{\Phi}=\Phi/\langle
\Phi\rangle$).}
\end{center}
\end{table}

\begin{table}
\begin{center}
\begin{tabular}{|c|r|r|}\hline
  & $\hat{\sigma}$\hspace{1cm} & $\hat{\var}$\hspace{1cm}  \\ \hline\hline
$ F_1 $ & $ -0.64090211.. $ & $ -3.70658437.. $ \\
$ F_2 $ & $  0.33867436.. $ & $  3.42228876.. $ \\
$ F_3 $ & $ -0.18662854.. $ & $ -2.38433446.. $ \\
$ F_4 $ & $  0.14277176.. $ & $  2.26840624.. $ \\
$ F_5 $ & $  0.06032607.. $ & $  1.21338371.. $ \\
$ F_6 $ & $ -0.04338937.. $ & $ -0.96176431.. $ \\
$ F_7 $ & $  0.01642569.. $ & $  0.45230320.. $ \\
$ F_8 $ & $ -0.00303607.. $ &$  -0.10584899.. $ \\ \hline
\end{tabular}
\caption{The one-particle form factors of the primary operators $\hat{\sigma}$ 
and $\hat{\varepsilon}$ ($\hat{\Phi}=\Phi/\langle\Phi\rangle$).}
\end{center}
\end{table}

Having determined the initial conditions of the form factor bootstrap for the
two relevant operators, all their form factors can in principle be computed
using the residue equations on dynamical and kinematical poles. The results for
several two-particle form factors are given in Table~2; Table~3 contains the 
full list of one-particle matrix elements \cite{immf,DS}. Referring to the 
normalisation-independent ratios (\ref{phihat}), all these numbers are 
universal. One can check that the factorisation conditions prescribed by 
(\ref{cluster}) are satisfied. Of course, the bootstrap can be continued to
determine the remaining two-particle form factors as well as those with more
than two particles (for example, $F^\sigma_{111}(\theta_1,\theta_2,\theta_3)$ 
is computed in \cite{immf}).

\vspace{.3cm}
The content of Tables~2 and 3 is sufficient to develop the large distance 
expansion (\ref{spectral}) of two-point correlators including all terms of
order lower than $e^{-3m_1|x|}$. In particular, the first few terms are
\EQ
\langle\hat{\Phi}_k(x)\hat{\Phi}_j(0)\rangle=1+
\frac1\pi\sum_{a=1}^3F_a^{\hat{\Phi}_k}F_a^{\hat{\Phi}_j}K_0(m_a|x|)+
O(e^{-2m_1|x|})\,,
\label{leading}
\EN
where $K_0(z)$ is a Bessel function.

The sum rules (\ref{cth}) and (\ref{Xth}) can be used to test the convergence
of the form factor expansion. We recall that $\Theta\sim\sigma$ and that the 
normalisation of the form factors of $\Theta$ is fixed by the condition 
(\ref{thetanorm}). The asymptotics
\bea
&& \langle\sigma(x)\sigma(0)\rangle\simeq\frac{C_{\sigma\sigma}^I}{|x|^{1/4}}
\nonumber\\
&& \langle\sigma(x)\varepsilon(0)\rangle\simeq
\frac{C_{\sigma\varepsilon}^\sigma}{|x|}\,\langle\sigma\rangle
\label{shortdistance}\\
&& \langle\varepsilon(x)\varepsilon(0)\rangle\simeq\frac{C_{\varepsilon
\varepsilon}^I}{|x|^2}\nonumber
\eea
as $|x|\goto 0$ follow from (\ref{ope}) and (\ref{isingope}) and are helpful 
in the interpretation of the results of Table~4 and 5 for the sum rules. The
faster convergence of the central charge sum rule is expected since the 
integration in $|x|^3d|x|$ strongly suppresses the contribution of short 
distances. In the scaling dimension sum rule the suppression is lower by a 
factor $|x|^2$ and more effective for $X_\sigma$ than for $X_\varepsilon$ since
$\langle\sigma\sigma\rangle$ is less singular than $\langle\sigma\varepsilon
\rangle$.

\begin{table}
\begin{center}
\begin{tabular}{|l|c|}\hline
$C_1 $ & 0.472038282  \\
$C_2 $ & 0.019231268 \\
$C_3 $ & 0.002557246 \\
$C_{11}$ & 0.003919717 \\
$C_4 $ & 0.000700348 \\
$C_{12}$ & 0.000974265 \\
$C_5 $ & 0.000054754 \\
$C_{13}$ & 0.000154186 \\ \hline
$C_{\rm partial}$ & 0.499630066 \\ \hline
\end{tabular}
\caption{Central charge from a partial sum of the form factor expansion of the 
correlation function in the sum rule (\ref{cth}). $C_{ab..}$ denotes the 
contribution of the state $A_aA_b..\,\,$. The exact result is $C=1/2$.}
\end{center}
\end{table}

\begin{table}
\begin{center}
\begin{tabular}{|l|c|c|}\hline
            &  $\sigma$   &  $\var$  \\ \hline
$\Delta_1 $ &   0.0507107 &   0.2932796 \\                                   
$\Delta_2 $ &   0.0054088 &   0.0546562 \\
$\Delta_3 $ &   0.0010868 &   0.0138858 \\
$\Delta_{11}$ & 0.0025274 &   0.0425125 \\
$\Delta_4 $ &   0.0004351 &   0.0069134 \\
$\Delta_{12}$ & 0.0010446 &   0.0245129 \\
$\Delta_5 $ &   0.0000514 &   0.0010340 \\ 
$\Delta_{13}$ & 0.0002283 &   0.0065067 \\ \hline
$\Delta_{\rm partial}$ & 0.0614934 & 0.4433015 \\ \hline
\end{tabular}
\caption{Conformal dimensions $\Delta_\Phi=X_\Phi/2$ from a partial sum of the 
form factor expansion of the correlation functions in the sum rule (\ref{Xth}).
$\Delta_{ab..}$ denotes the contribution of the state $A_aA_b..\,\,$. The exact
results are $\Delta_\sigma=1/16=0.0625$ and $\Delta_\varepsilon=1/2$.}
\end{center}
\end{table}

It is of obvious interest to compare the results of integrable field theory
discussed in this section with the numerical results for the original lattice 
model (\ref{lattice}) with $T=T_c$. Actually, the first numerical 
investigation was performed in \cite{HS} on the Ising quantum spin chain with 
the purpose of confirming the Zamolodchikov's mass spectrum, what was done with
good precision for the first few masses\footnote{The full spectrum has been 
recovered in \cite{BNW} from the exact solution of the RSOS model of 
\cite{WNS}, which is in the same universality class of the magnetic Ising 
model. The spectrum of the field theory (\ref{A}) with $\tau=0$ was also 
studied in \cite{SZ} by numerical diagonalisation of the Hamiltonian on the 
(suitably truncated) space of states of the conformal theory 
\cite{YZtruncation}.}. Concerning correlation functions, after the early 
studies of \cite{LR,DdRORT}, the most accurate Monte Carlo simulations have 
been performed in \cite{CGM}. In this work the numerical data have been used
to test the form factor expansion as well as the corrections to the short 
distance behaviour computed in \cite{GM}; the analysis has confirmed the good 
convergency properties of both expansions.

Caselle and Hasenbusch used a numerical diagonalisation of the transfer matrix
of the lattice model to evaluate the first few one-particle form factors 
\cite{CH1}. The comparison of their results (Table~6) with those of Table~3 
provides a direct check of the form factor computations reviewed in this 
section. The transfer matrix method has also provided a test of two-particle 
form factors. Indeed, one-point functions on a cylinder of 
circumference $R$ behave as
\EQ
\frac{\langle\Phi\rangle_R}{\langle\Phi\rangle_{R=\infty}}=
1+\frac{1}{\pi}\sum_a\,A^\Phi_a\,K_0(m_aR)+O(e^{-2m_1R})\,,\hspace{1cm}
R\goto\infty
\EN
where the amplitudes
\EQ
A^\Phi_a=\left.\frac{F_{aa}^\Phi(i\pi)}{\langle\Phi\rangle}
\right|_{R=\infty}
\EN
are given in Table~7. The avalaible numerical estimates are \cite{CH2} 
\EQ
A_1^\sigma=-8.11(2)\,,\hspace{1.5cm}A_1^\var=-17.5(5)\,\,.
\EN

\begin{table}
\begin{center}
\begin{tabular}{|c|l|l|}\hline
& $\hspace{.7cm}\hat{\sigma}$ & $\hspace{.6cm}\hat{\var}$  \\ \hline\hline
$ |F_1| $ & $ 0.6408(3) $ & $ 3.707(7) $ \\
$ |F_2| $ & $ 0.325(25) $ & $ 3.38(7) $ \\ \hline
\end{tabular}
\caption{Lattice estimates of one-particle form factors obtained in 
\cite{CH1}.} 
\end{center}
\end{table}

\begin{table}
\begin{center}
\begin{tabular}{|c|c|c|}\hline
               & $\sigma$ & $\varepsilon$ \\ \hline\hline
$ A_1$ & $-8.0999744..$ & $-17.893304..$      \\
$ A_2$ & $-21.206008..$ & $-24.946727..$      \\
$ A_3$ & $-32.045891..$ & $-53.679951..$      \\ \hline
\end{tabular}
\caption{Universal amplitudes ruling the leading finite size corrections of 
one-point functions \cite{ft}.}
\end{center}
\end{table}

\resection{Ising universality class}
Two statistical mechanical systems may differ for a number of microscopic 
features (e.g the type of lattice or the number of neighbouring sites entering 
the interaction term) and still be characterised by the same internal symmetry.
In this case, the systems will exhibit the same critical behaviour nearby a
second order phase transition point associated to the spontaneous breaking of
the symmetry. As it is said, they belong to the same universality class. 
Quantum field theory deals directly with the continuum limit in which the 
microscopic details become immaterial, and is the natural framework for 
describing universality classes. In particular, the action (\ref{A}) describes 
the universality class of the two-dimensional Ising model, and all the results
discussed so far with reference to this action are universal. Traditionally, 
however, the quantitative characterisation of universality in 
statistical mechanics is made studying the behaviour of the thermodynamical 
observables in the vicinity of the critical point. In this section we show how 
integrable field theory allows to complete the list of canonical universal
quantities of the Ising universality class \cite{ratios}. 

The usual procedure (see \cite{PHA} and references therein) is that of 
introducing critical exponents and critical amplitudes through the relations 
\bea
\xi &=& \left\{\matrix{
    f_\pm\,|\tau|^{-\nu},\hspace{1cm}\tau\rightarrow 0^\pm,
      \hspace{.2cm}h=0 \cr\cr
    f_c\,|h|^{-\nu_c},\hspace{1cm}\tau=0, 
      \hspace{.2cm}h\rightarrow 0 \cr}\right.
\nonumber\\\nonumber\\\nonumber\\
C &=& \left\{\matrix{
   (A_\pm/\alpha)\,|\tau|^{-\alpha},\hspace{1cm}\tau\rightarrow 0^\pm,
\hspace{.2cm}h=0 \cr\cr
(A_c/\alpha_c)\,|h|^{-\alpha_c},\hspace{1cm}\tau=0, 
\hspace{.2cm}h\rightarrow 0 \cr}\right.
\nonumber\\\nonumber\\\nonumber\\
|{\cal M}| &=& \left\{\matrix{
   B\,(-\tau)^{\beta},\hspace{1cm}\tau\rightarrow 0^-,
\hspace{.2cm}h=0 \cr\cr
(|h|/D)^{1/\delta},\hspace{1cm}\tau=0, 
\hspace{.2cm}h\rightarrow 0 \cr}\right.
\nonumber\\\nonumber\\\nonumber\\
\chi &=& \left\{\matrix{
    \Gamma_\pm\,|\tau|^{-\gamma},\hspace{1cm}\tau\rightarrow 0^\pm,
      \hspace{.2cm}h=0 \cr\cr
    \Gamma_c\,|h|^{-\gamma_c},\hspace{1cm}\tau=0, 
      \hspace{.2cm}h\rightarrow 0 \cr}\right.
\nonumber
\eea
where $\xi$ is the correlation length, $C$ the specific heat, ${\cal M}$ the 
magnetisation and $\chi$ the susceptibility; $\tau$ and $h$ are the deviation
from critical temperature and the magnetic field entering the action (\ref{A}).
The observables above are related to the one- and two-point functions of the
scaling operators $\sigma$ and $\varepsilon$. After introducing the free
energy per unit area
\EQ
f=-\frac{1}{A}\ln Z\,,
\EN
we have\footnote{We recall that $\langle\cdots\rangle_c$ denotes connected
correlators.}
\bea
C &=& -\frac{\partial^2 f}{\partial\tau^2}=\int d^2x\,\langle\var(x)\var(0)
\rangle_{c}
\label{heat}\\
{\cal M} &=&-\frac{\partial f}{\partial h}=\langle\sigma\rangle\\
\chi &=&-\frac{\partial^2 f}{\partial h^2}=\int d^2x\,\langle\sigma(x)\sigma(0)
\rangle_{c}\,\,.
\eea
As for the correlation length, it is common to distinguish between the `true' 
correlation length $\xi_t$ defined as
\EQ
\lim_{|x|\goto\infty}\langle\sigma(x)\sigma(0)\rangle_c\sim e^{-|x|/\xi_t}\,,
\label{xitrue}
\EN
and the second moment correlation length
\EQ
\xi_{2nd}=\left(\frac{1}{4\chi}\int d^2x\,|x|^2\langle\sigma(x)\sigma(0)
\rangle_c\right)^{1/2}\,\,.
\label{xi2nd}
\EN
We will distinguish the critical amplitudes for the two types of correlation
length through the superscripts $t$ and $2nd$.

If $M\sim 1/\xi$ is a mass scale, we know that $f\sim M^2$, $\tau\sim
M^{2-X_\varepsilon}$ and $h\sim M^{2-X_\sigma}$. Then, comparison of 
(\ref{heat})-(\ref{xi2nd}) with the definitions of the critical exponents 
gives
\bea
&& \nu=1/(2-X_\var)=1\nonumber\\
&& \nu_c=1/(2-X_\sigma)=8/15\nonumber\\
&& \alpha=2(1-X_\var)\,\nu=0\nonumber\\
&& \alpha_c=2(1-X_\var)\,\nu_c=0\nonumber\\
&& \beta=X_\sigma\,\nu=1/8\nonumber\\
&& \delta=1/(X_\sigma\,\nu_c)=15\nonumber\\
&& \gamma=2(1-X_\sigma)\,\nu=7/4\nonumber\\
&& \gamma_c=2(1-X_\sigma)\,\nu_c=14/15\,\,.
\nonumber
\eea
These well known results express the fact that the critical exponents are 
universal and determined by the scaling dimensions of the operators $\sigma$
and $\varepsilon$. They also account for the usual scaling and hyperscaling
relations
\bea
&& \alpha+2\beta+\gamma=2\label{scal1}\\
&& \alpha+2\nu=2\\
&& \gamma=\beta(\delta-1)\\
&& \alpha_c=\alpha/\beta\delta\\
&& \nu_c=\nu/\beta\delta\label{scal5}\\
&& \gamma_c=1-1/\delta\,\,.\label{scal6}
\eea

Contrary to the exponents, the critical amplitudes are not determined by 
conformal field theory and are not themselves universal. Roughly speaking, 
they depend on the scales we use to measure the temperature and the magnetic 
field (metric factors). It is possible, however, to combine different 
amplitudes in such a way that the metric factors cancel leaving universal 
quantities. Such universal amplitude combinations provide the canonical 
characterisation of the universality of the scaling region surrounding the 
critical point.

A simple way of canceling the metric factors is to consider ratios like
$\Gamma_+/\Gamma_-$, $f^{2nd}_c/f^t_c$ and similar. More sophisticated
cancelations are obtained exploiting the relations (\ref{scal1})--(\ref{scal6})
among the critical exponents. So, to (\ref{scal1})--(\ref{scal5}) one 
associates the universal quantities \cite{PHA}
\bea
&& R_c=A_+\Gamma_+/B^2\\
&& R_\xi^+=A^{1/2}_+\,f_+^t\\
&& R_\chi=\Gamma_+DB^{\delta-1}\\
&& R_A=A_cD^{-(1+\alpha_c)}B^{-2/\beta}\\
&& Q_2=(\Gamma_+/\Gamma_c)(f_c^t/f_+^t)^{\gamma/\nu}\,\,.
\eea
Equation (\ref{scal6}) is instead associated to the identity
\EQ
\delta\Gamma_cD^{1/\delta}=1
\label{gammadelta}
\EN
which can be recovered from (\ref{Xth}) with $\Phi=\sigma$ and
\EQ
\Theta=-2\pi h(2-X_\sigma)\,\sigma\,,\hspace{1cm}\tau=0\,\,.
\label{Thetasigma}
\EN

Notice that the critical amplitudes of the Ising universality class are 
defined along the integrable directions of the field theory (\ref{A}). As we 
are going to see, all of them can be computed exactly. We will work within the 
so-called `conformal normalisation' of the operators $\sigma$ and 
$\varepsilon$, which amounts to take 
\EQ
C_{\sigma\sigma}^I=C_{\varepsilon\varepsilon}^I=1
\label{cftnorm}
\EN
in (\ref{shortdistance}). We know that the mass $m$ of the particle at $h=0$
and the mass $m_1$ of the lightest particle at $\tau=0$ can be written as
\bea
&& m={\cal C}_\tau|\tau|\label{thermalmass}\\
&& m_1={\cal C}_h|h|^{8/15}\label{magneticmass}\,\,.
\eea
The two constants are known exactly from a modified version of the 
thermodynamic Bethe ansatz, and read \cite{SGmass,Fateev}
\bea
&& {\cal C}_\tau=2\pi\\
&& {\cal C}_h=4.40490857..\,\,.
\eea

The spontaneous magnetisation at $h=0$ was given in (\ref{spontaneous}) and
corresponds to the amplitude
\EQ
B=1.70852190..\,\,.
\EN
In the magnetic direction, we can use (\ref{cluster}) and (\ref{Thetasigma}) 
to write
\EQ
\langle\sigma\rangle=[-2\pi h(2-X_\sigma)]^{-1}\frac{(F_1^\Theta)^2}
{\lim_{\theta\goto\infty}F_{11}^\Theta(\theta)}\,,\hspace{1cm}\tau=0
\EN
where the form factors of $\Theta$ coincide with those determined in section~4 
for $\hat{\sigma}$, up to the normalisation fixed by (\ref{thetanorm}). The 
result for the magnetisation obtained in this way concides with that given by
the thermodynamic Bethe ansatz \cite{Fateev}, and provides the amplitude
\EQ
D=0.0253610264..\,\,.
\label{magampl}
\EN

The high- and low-temperature susceptibilities at $h=0$ are obtained 
integrating the correlators (\ref{sigmasigma}) and (\ref{mumu}) with the 
normalisation (\ref{spontaneous}), what gives
\bea
&& \Gamma_+=0.148001214..\\
&& \Gamma_-=0.00392642280..\,\,.
\eea
The susceptibility amplitude at $\tau=0$ follows from (\ref{gammadelta}) and 
(\ref{magampl}) and reads
\EQ
\Gamma_c=0.0851721517..\,\,.
\EN

It appears from (\ref{xitrue}) that the `true' correlation lenght is just the 
inverse of the total mass of the lightest particle state entering the form
factor expansion of the spin-spin correlator. Then we know from the analysis 
of section~4 that 
\bea
&& \xi_t=\left\{\matrix{1/m\,,\hspace{1cm}
\tau>0\cr\cr
                      1/2m\,,\hspace{.8cm}\tau<0\cr}\right.
\,,\hspace{2cm}h=0\\\nonumber\\
&& \xi_t=1/m_1\,,\hspace{4.5cm}\tau=0\,\,.
\eea
Concerning the second moment correlation length, at $h=0$ the integration of 
the exact correlators gives
\EQ
\xi_{2nd}=\left\{\matrix{(0.99959808..)m\,,\hspace{1cm}
\tau>0\cr\cr
                         (0.31607894..)m\,,\hspace{1cm}\tau<0\cr}\right.
\,,\hspace{2cm}h=0\,\,.
\EN
At $\tau=0$ we do not know the exact spin-spin correlator, but the integral
entering (\ref{xi2nd}) is related to the central charge sum rule (\ref{cth}) 
by (\ref{Thetasigma}). Putting all together one finds the amplitude
\EQ
f_c^{2nd}=\sqrt{\frac{8}{45\pi}}\,D^{1/30}=0.21045990..\,\,.
\EN

Dealing finally with the specific heat, we have to take into account that, 
since $\alpha=\alpha_c=0$, the critical behaviour gets modified into
\EQ
C\simeq\left\{\matrix{-A_\pm\ln|\tau|\,,\hspace{1cm}h=0\cr\cr
                 -A_c\ln|h|\,,\hspace{1cm}\tau=0\,\,.}\right.
\EN
On the other hand, (\ref{shortdistance}) and (\ref{cftnorm}) allow to isolate
the logarithmic singularity as 
\EQ
C=\int d^2x\,\langle\var(x)\var(0)\rangle_c\sim 2\pi\int_{r_0}\frac{dr}{r}
\sim -2\pi\ln(Mr_0)\,,
\EN
where $r=|x|$, $r_0$ is a short distance cut-off and $M$ is a mass scale 
needed to make dimensionless the argument of the logarithm. It will be 
propotional to $m$ at $h=0$ and to $m_1$ at $\tau=0$. Recalling 
(\ref{thermalmass}) and (\ref{magneticmass}) we conclude
\bea
&& A_\pm=2\pi\\
&& A_c=\frac{16\pi}{15}\,\,.
\eea

The values of the universal amplitude combinations which follow from these 
results are collected in Table~8. The first seven quantities involve only 
amplitudes computed at $h=0$ and are all known since the work of \cite{WMcTB};
the last four numbers involve also amplitudes computed along the magnetic
direction and where given in \cite{ratios}. For the latter, an early lattice 
estimate coming from series expansions is $R_\chi\sim 6.78$ \cite{TF}; more 
recently, transfer matrix techniques have given the results $R_\chi=6.7782(8)$ 
and $Q_2=3.233(4)$ \cite{CH1}.

\begin{table}
\begin{center}
\begin{tabular}{|c|}\hline
$A_+/A_-=1$  \\
$\Gamma_+/\Gamma_-=37.6936520.. $\\
$f^t_+/f^t_-=2$ \\
$f^{2nd}_+/f^t_+=0.99959808..$ \\
$f^{2nd}_+/f^{2nd}_-=3.16249504..$ \\
$R_C=0.318569391..$ \\
$R_\xi^+=1/\sqrt{2\pi}=0.39894228..$ \\
$f^{2nd}_c/f^t_c=0.9270566..$ \\
$R_\chi=6.77828502..$  \\
$R_A=0.0250658794..$ \\
$Q_2=3.23513834..$  \\
\hline
\end{tabular}
\caption{The universal amplitude combinations of the Ising universality class.}
\end{center}
\end{table}

\resection{Beyond integrability}
We saw in section~3 how two-dimensional quantum field theories allow for 
integrable directions in coupling space along which a great deal of exact 
information can be obtained. A number of consequences of this fact for the 
scaling Ising model have been discussed in the last two sections. It is quite 
natural to ask how the exact results for the integrable directions can be 
exploited
for the analytic study of the non-integrable ones \cite{nonint}. In general, 
we can decompose the action ${\cal A}_{NI}$ of the non-integrable theory into 
the action ${\cal A}_I$ describing an integrable direction plus the 
contributions of the scaling operators responsible for the deviation from 
integrability. Considering, for the sake of simplicity, the subspace spanned 
by only one of such operators, we write
\EQ
{\cal A}_{NI}={\cal A}_{I}-\lambda\int d^2x\,\Psi(x)\,\,.
\label{nonint}
\EN
In principle, this theory can be studied by perturbation theory in $\lambda$:
all the corrections can be formally written in terms of the matrix elements
of the perturbing operator $\Psi$ between particle states in the unperturbed
(integrable) theory; as we know, these matrix elements can be computed exactly.

The first order corrections are particularly simple and normally sufficient
to explore small deviations from integrability. For example, the leading 
corrections to the energy spectrum and the scattering amplitudes for the case 
of neutral particles read
\cite{nonint}
\bea
& & \delta{\cal E}_{vac}=-\left.\langle\Psi\rangle\right|_{\lambda=0}\,\lambda
+{\cal O}(\lambda^2) \\
& & \delta m^2_a=-2\left.F^\Psi_{aa}(i\pi)\right|_{\lambda=0}\,\lambda
+{\cal O}(\lambda^2)\label{deltam}\\
&& \delta S_{ab}(\theta)=i\left.\frac{F^\Psi_{abab}(\theta)}{m_am_b}
\right|_{\lambda=0}\,\frac{\lambda}{\sinh\theta}+{\cal O}(\lambda^2)\,,
\eea
where
${\cal E}_{vac}$ is the energy density of the vacuum state and\footnote{We
denote by $\langle\cdots|\cdots|\cdots\rangle_{c}$ the connected part of a 
matrix element; $\Psi$ is a scalar operator.}
\EQ
F^\Psi_{aa}(i\pi)\equiv\langle A_a(\theta)|\Psi(0)|A_a(\theta)
\rangle_{c}=F^\Psi_{aa}(\theta+i\pi,\theta)
\EN
\EQ
F^\Psi_{abab}(\theta_1-\theta_2)\equiv\langle A_a(\theta_1)A_b(\theta_2)|
\Psi(0)|A_a(\theta_1)A_b(\theta_2)\rangle_{c}=F^\Psi_{abab}(\theta_1+
i\pi,\theta_2+i\pi,\theta_1,\theta_2)\,\,.
\EN

Going to the Ising field theory (\ref{A}), one possibility is to identify the 
magnetic direction\footnote{The parameter $\eta$ which labels the 
renormalisation group trajectories in the plane of Figure~1 was defined in 
(\ref{eta}).}
$\eta=0$ as the integrable unperturbed theory, and to take $\lambda\Psi=
\tau\varepsilon$ as the perturbation. Then, in particular, the ratios
\EQ
\frac{\delta m_a^2}{\delta{\cal E}_{vac}}=2\left.F^{\hat{
\varepsilon}}_{aa}(i\pi)\right|_{\eta=0}+{\cal O}(\eta)\,,\hspace{1cm}
\eta\goto 0
\label{deltama}
\EN
are completely universal and can be checked in any numerical approach
giving access to the energy spectrum\footnote{When the system is compactified 
on a cylinder of circumference $R$, the ground state energy behaves as 
${\cal E}_{vac}R$ for $R/\xi\gg 1$.}. Their values for $a=1,2,3,4$ follows
from the results of Table~2, which give
\bea
&& \left.F^{\hat{\varepsilon}}_{11}(i\pi)\right|_{\eta=0}=-17.8933..\\
&& \left.F^{\hat{\varepsilon}}_{22}(i\pi)\right|_{\eta=0}=-24.9467..\\
&& \left.F^{\hat{\varepsilon}}_{33}(i\pi)\right|_{\eta=0}=-53.6799..\\
&& \left.F^{\hat{\varepsilon}}_{44}(i\pi)\right|_{\eta=0}=-49.3169..\,\,.
\eea
The first few ratios (\ref{deltama}) have been checked with good accuracy
by numerical diagonalisation of the Hamiltonian of the Ising field theory on a 
conformal basis of states \cite{nonint}, and in the lattice model by numerical 
diagonalisation of the transfer matrix \cite{GR}. The absolute variations of
the energy levels can be obtained using the result
\EQ
\left.\langle\varepsilon\rangle\right|_{\eta=0}=(2.00314..)\,|h|^{8/15}
\EN
which refers to the conformal normalisation (\ref{cftnorm}) of the operators 
and has been obtained in \cite{vevs} through quite sophisticated techniques;
a previous lattice estimate \cite{GMvevs} agrees well with this exact 
value.

Notice that $\delta m_a$ will present an imaginary contribution at order 
$\tau^2$ for $a=4,\ldots,8$. Indeed, we know that at $\tau=0$ the masses
$m_a$ with $a>3$ lie above the lowest two-particle threshold $2m_1$. There is
nothing wrong with this as long as the theory is integrable and all inelastic
processes are forbidden. Moving away from $\tau=0$, however, integrability
is lost and nothing can prevent the decay $A_a\goto A_1A_1$ when $a>3$. The 
imaginary part in the mass is inversely proportional to the lifetime of the 
unstable particle. For $A_4$ it is given by \cite{nonint}
\EQ
\mbox{Im}\,m_4^2\simeq -\left.\frac{|F^\varepsilon_{411}(i\pi,\theta^*,
-\theta^*)|^2}{m_1m_4}\right|_{\eta=0}\,\frac{\tau^2}{2\sinh\theta^*}\,,
\hspace{1cm}
\EN
with $\theta^*$ determined by energy conservation; heavier particles  
receive similar contributions by other decay channels. These decay widths, as 
well as the corrections to the scattering amplitudes, can in principle be 
computed turning the crank of the form factor bootstrap.

In a similar way, the region close to the thermal axis in Figure~1 can be 
analysed perturbing the $h=0$ theory with the operator $\Psi=\sigma$. Let us 
start from the low-temperature phase ($\eta=-\infty$). We know from 
(\ref{deltam}) 
and (\ref{mueven}) that in this case the first order correction to the mass of 
the particle of the unperturbed theory is proportional to $\tanh(i\pi/2)$, and 
then is infinite. This result is easily understood once we remember that the
particle we are dealing with is the kink interpolating between the two vacua of
the spontaneously broken phase at $h=0$. As soon as we switch on the magnetic 
field, the degeneracy of the vacua is lifted, so that all states carrying 
a topologic charge acquire an infinite energy and are removed from the theory;
the single particle excitations of the perturbed theory are kink-antikink 
bound states $A_n$. Hence, the divergence of the mass correction is the 
signature of the {\em confinement} of the topologic charge and manifests itself
whenever the perturbing operator $\Psi$ is mutually non-local with respect to 
the particles of the unperturbed theory\footnote{We recall that in the present 
case the kink is created by the disorder operator $\mu$.} \cite{nonint,msg};
indeed, (\ref{ff4}) shows that this is the condition for having a pole at 
$\theta=i\pi$ in $F^\Psi_{aa}(\theta)$.

The magnetic field induces an energy density difference
\EQ
2|\delta{\cal E}_{vac}|=2vh+{\cal O}(h^2)
\EN
between the two vacua, $v$ being the spontaneous magnetisation 
(\ref{spontaneous}) of the theory at $\eta=-\infty$ ($v$ takes the sign of 
$h$). Then, neglecting the relativistic effects, a kink
and an antikink separated by a distance $R$ are bounded by the potential
$2vhR$. The associated Schrodinger problem gives for the bound states $A_n$ 
the masses
\EQ
m_n\simeq 2m+\frac{(2vh)^{2/3}z_n}{m^{1/3}}\,,
\label{nonrel}
\EN
where $m$ is the mass of the kink and $z_n$, $n=1,2,\ldots$, are positive
numbers determined by the zeroes of the Airy function, Ai$(-z_n)=0$. This 
non-relativistic approximation holds as long as $m_n-2m\ll m$, and then gives 
the correct leading behaviour for $\eta\goto -\infty$. Of course, the 
particles with masses $m_n$ larger than twice the
lightest mass $m_1$ are unstable. The masses $m_n$ become dense 
above $2m$ as $\eta\goto -\infty$, and condensate to reproduce the
continuum spectrum when $h$ vanishes. The spectrum (\ref{nonrel}) was first
obtained in \cite{McW} through the study of the analytic structure in momentum
space of the spin-spin correlation function for small magnetic field. 
Relativistic corrections to (\ref{nonrel}) have recently been obtained in
\cite{FZ1,FZ2}.

It is possible at this point to figure out the evolution of the
particle spectrum of the Ising field theory in the plane of Figure~1. 
We have just described how the kinks living 
at $\eta=-\infty$ are confined into a tower of bound states when the magnetic 
field is switched on. The number of stable particles (i.e. with $m_n<2m_1$) 
decreases as $\eta$ is increased from $-\infty$ to finite negative values.
We expect three such particles to be left for small values of $\eta$ (i.e.
close to the magnetic axis), although five more particles above threshold 
happen to be stable at $\eta=0$ due to integrability. The number of stable 
particles continues to decrease as $\eta$ increases, until a single particle is
left at $\eta=+\infty$ (the positive thermal axis). This scenario, which is an 
updated version of that originally proposed in \cite{McW}, is supported by the 
numerical investigation of the spectrum of the Ising field theory 
\cite{nonint,FZ1}.

The correction to the mass of the particle when the magnetic field is 
switched on at $\tau>0$ can again be computed in perturbation theory. This 
time, however, the first order correction vanishes for symmetry reasons (the 
perturbing operator $\sigma$ couples only to an odd number of particles at 
$\eta=+\infty$), and one has
\EQ
\delta m_1^2=b\,m^{-7/4}h^2+{\cal O}(h^4)\,,
\EN
where $m$ is the mass at $\eta=+\infty$, and 
\EQ
b=-m^{7/4}\,\int d^2x\,\langle A(\theta)|\sigma(x)\sigma(0)|A(\theta)
\rangle_{c}
\label{2ndorder}
\EN
(the integral is evaluated at $\eta=+\infty$). In principle, one could think 
to evaluate the matrix element through a decomposition over intermediate 
particle states. Very recently, however, Fonseca and A.~Zamolodchikov computed 
it generalising the techniques used for the spin-spin correlation function at 
$h=0$. In this way they found \cite{FZ2}
\EQ
b=21.52398..
\EN
within the normalisation (\ref{cftnorm}) of the operators. Since
\EQ
\delta{\cal E}_{vac}=-\frac{h^2}{2}\int d^2x\,\langle\sigma(x)\sigma(0)
\rangle_c+{\cal O}(h^4)=-\frac{\Gamma_+}{2}\,\tau^{-7/4}h^2+{\cal O}(h^4)\,,
\EN
we have
\EQ
\frac{\delta m_1^2}{\delta{\cal E}_{vac}}=-\frac{2b}{\Gamma_+}\,(2\pi)^{-7/4}
+{\cal O}(\eta^{-15/4})=-(11.66467..)+{\cal O}(\eta^{-15/4})\,,\hspace{1cm}
\eta\goto +\infty\,\,.
\EN

We mention, in conclusion, that \cite{FZ1} contains also a detailed study,
which combines analytic expansions and numerical methods, of the analytic 
properties of the free energy of the Ising field theory as a function of 
temperature and magnetic field.

\vspace{1cm}
{\bf Acknowledgments.}~~I thank J.~Cardy, G.~Mussardo and P.~Simonetti for
collaborating with me on several topics reviewed in this article. This work
was partially supported by the European Commission TMR programme 
HPRN-CT-2002-00325 (EUCLID) and by the COFIN ``Teoria dei Campi, Meccanica
Statistica e Sistemi Elettronici''.


\newpage


\begin{thebibliography}{99}

\bibitem{Ising} E. Ising, Z. Physik 31 (1925) 253.
\bibitem{Onsager} L. Onsager, Phys. Rev. 65 (1944) 117.
\bibitem{Yang} C.N. Yang, Phys. Rev. 85 (1952) 808.
\bibitem{McCoyWu} B.M. McCoy and T.T. Wu, The Two-Dimensional Ising Model, 
Harvard University Press, Cambridge, Massachussets, 1973.
\bibitem{WMcTB}T.T. Wu, B.M. Mccoy, C.A. Tracy and E. Barouch, Phys. Rev. B 13 
(1976) 316.
\bibitem{Taniguchi} A.B. Zamolodchikov, Advanced Studies in Pure
Mathematics 19 (1989) 641; Int. J. Mod. Phys. A 3 (1988) 743.
\bibitem{BPZ} A.A. Belavin, A.M. Polyakov and A.B. Zamolodchikov, Nucl. Phys.
B 241 (1984) 333.
\bibitem{WNS} S.O. Waarnar, B. Nienhuis and K.A. Seaton, Phys. Rev. Lett. 69 
(1992) 710.
\bibitem{immf} G. Delfino and G. Mussardo, Nucl. Phys. B 455 (1995) 724.
\bibitem{DS} G. Delfino and P. Simonetti, Phys. Lett. B 383 (1996) 450.
\bibitem{KWduality} H.A. Kramers and G.H. Wannier, Phys. Rev. 60 (1941) 252
and 263.
\bibitem{Cardybook} J.L. Cardy, Scaling and renormalization in statistical
physics, Cambridge University Press, Cambridge, 1996.
\bibitem{FQS} D. Friedan, Z. Qiu and S. Shenker, Phys. Rev. Lett. 52 (1984) 
1575.
\bibitem{KC} L.P. Kadanoff and H. Ceva, Phys. Rev. B 3 (1971) 3918.
\bibitem{Giuseppe} G. Mussardo, Phys. Rep. 218 (1992) 215.
\bibitem{Fonseca} P.D. Fonseca, Mod. Phys. Lett. A 13 (1998) 1931.
\bibitem{ZZ} A.B. Zamolodchikov and Al.B. Zamolodchikov, Ann. Phys. 120 (1979) 
253.
\bibitem{SW} R. Shankar and E. Witten, Phys. Rev. D 17 (1978) 2134.
\bibitem{Iagolnitzer} D. Iagolnitzer, Phys. Lett. B 76 (1978) 207;
Phys. Rev. D 18 (1978) 1275.
\bibitem{Parke} S. Parke, Nucl. Phys. B 174 (1980) 166.
\bibitem{ELOP} R.J. Eden, P.V. Landshoff, D.I. Olive and J.C. Polkinghorne,
The analytic S-matrix, Cambridge University Press, 1966.
\bibitem{Mitra} P. Mitra, Phys. Lett. B 72 (1977) 62.
\bibitem{KW} M. Karowski, P. Weisz, Nucl. Phys. B 139 (1978) 455.
\bibitem{Smirnov} F.A. Smirnov, Form Factors in Completely Integrable
Models of Quantum Field Theory, World Scientific, 1992.
\bibitem{YZ} V.P. Yurov and Al.B. Zamolodchikov, Int. J. Mod. Phys.
A 6 (1991) 3419.
\bibitem{AlyoshaYL} Al.B. Zamolodchikov, Nucl. Phys. B 348 (1991) 619. 
\bibitem{Zamocth} A.B. Zamolodchikov, JETP Lett. 43 (1986) 730.
\bibitem{Cardycth} J.L. Cardy, Phys. Rev. Lett. 60 (1988) 2709.
\bibitem{DSC} G. Delfino, P. Simonetti and J.L. Cardy, Phys. Lett. B 387 
(1996) 327. 
\bibitem{BKW} B. Berg, M. Karowski, P. Weisz, Phys. Rev. D 19 (1979) 2477. 
\bibitem{CM} J.L. Cardy and G. Mussardo, Nucl. Phys. B 340 (1990) 387.
\bibitem{BB} O. Babelon and D. Bernard, Phys. Lett. B 288 (1992) 113.
\bibitem{BL} D. Bernard and A. LeClair, Nucl. Phys. B 426 (1994) 534; 
Erratum-ibid. B 498 (1997) 619.
\bibitem{SMJ} M. Sato, T. Miwa and T. Jimbo, Publ. RIMS Kyoto Univ. 14 (1978)
223.
\bibitem{FZ2} P. Fonseca and A.B. Zamolodchikov, hep-th/0309228.
\bibitem{BNNPSW} J. Balog, M. Niedermaier, F. Niedermayer, A. Patrascioiu,
E. Seiler and P. Weisz, Nucl. Phys. B 583 (2000) 614.
\bibitem{FateevE8} V.A. Fateev, unpublished.
\bibitem{FZE8} V.A. Fateev and A.B. Zamolodchikov, Int. J. Mod. Phys. A 5
(1990) 1025.
\bibitem{CT} S. Coleman and H.J. Thun, Commun. Math. Phys. 61 (1978) 31.
\bibitem{BCDS} H.W. Braden, E. Corrigan, P.E. Dorey and R. Sasaki, Nucl. Phys.
B 338 (1990) 689.
\bibitem{ChM} P. Christe and G. Mussardo, Nucl. Phys. B 330 (1990) 465.
\bibitem{TBA} Al.B. Zamolodchikov, Nucl. Phys. B 342 (1990) 695. 
\bibitem{KM} T.R. Klassen and E. Melzer, Nucl. Phys. B 338 (1990) 485.
\bibitem{HS} M. Henkel and H. Saleur, J. Phys. A 22 (1989) L513.
\bibitem{BNW} V.V. Bazhanov, B. Nienhuis, and S.O. Warnaar, Phys. Lett. B 322
(1994) 198.
\bibitem{SZ} I.R. Sagdeev and A.B. Zamolodchikov, Mod. Phys. Lett. B 3 (1989) 
1375.
\bibitem{YZtruncation} V.P. Yurov and Al.B. Zamolodchikov, Int. J. Mod. Phys.
A 6 (1991) 4557.
\bibitem{LR} P.G. Lauwers and V. Rittenberg, Phys. Lett. B 233 (1989) 197;
Bonn preprint HE-89-11. 
\bibitem{DdRORT} C. Destri, F. Di Renzo, E. Onofri, P. Rossi and G.P. 
Tecchiolli, Phys. Lett. B 278 (1992) 311.
\bibitem{CGM} M. Caselle, P. Grinza and N. Magnoli, Nucl. Phys. B 579 (2000) 
635.
\bibitem{GM} R. Guida and N. Magnoli, Nucl. Phys. B 483 (1997) 563.
\bibitem{CH1} M. Caselle and M. Hasenbusch, Nucl. Phys. B 579 (2000) 667.
\bibitem{CH2} M. Caselle and M. Hasenbusch, Nucl. Phys. B 639 (2002) 549.
\bibitem{ft} G. Delfino, J. Phys. A 34 (2001) L161.
\bibitem{ratios} G. Delfino, Phys. Lett. B 419 (1998) 291; Erratum-ibid. B 518
(2001) 330. 
\bibitem{PHA} V. Privman, P.C. Hohenberg and A. Aharony, Universal critical
point amplitude relations, in `Phase transition and critical phenomena',
Vol. 14, C. Domb and J.L. Lebowitz eds., Academic Press, 1991. 
\bibitem{SGmass} Al.B. Zamolodchikov, Int. J. Mod. Phys. A 10 (1995) 1125.
\bibitem{Fateev} V.A. Fateev, Phys. Lett. B 324 (1994) 45.
\bibitem{TF} H.B. Tarko and M.E. Fisher, Phys. Rev. B 11 (1975) 1217.
\bibitem{nonint} G. Delfino, G. Mussardo and P. Simonetti, Nucl. Phys. B 473 
(1996) 469.
\bibitem{GR} P. Grinza and A. Rago, Nucl. Phys. B 651 (2003) 387.
\bibitem{vevs} V.A. Fateev, S. Lukyanov, A.B. Zamolodchikov and 
Al.B.~Zamolodchikov, Nucl. Phys. B 516 (1998) 652.
\bibitem{GMvevs} R. Guida and N. Magnoli, Phys. Lett. B 411 (1997) 127.
\bibitem{msg} G. Delfino and G. Mussardo, Nucl. Phys. B 516 (1998) 675.
\bibitem{McW} B.M. McCoy and T.T. Wu, Phys. Rev. D 18 (1978) 1259.
\bibitem{FZ1} A.B. Zamolodchikov and P. Fonseca, J. Stat. Phys. 110 (2003) 527.


\end{thebibliography}
\end{document}